\documentclass[pre, reprint, sd, amsmath, amssymb, notitlepage, superscriptaddress]{revtex4-2}

\usepackage{amsmath}
\usepackage{amsfonts}
\usepackage{amssymb}
\usepackage{amsthm}
\usepackage{fancyhdr}
\usepackage{enumerate}
\usepackage{graphicx}
\usepackage[labelfont=bf]{caption}

\usepackage{color}

\usepackage{caption}
\usepackage{subcaption}

\usepackage{verbatim}

\usepackage{wasysym}
\usepackage{hyperref}

\usepackage{wrapfig}

\newcommand{\Exp}[1]{\langle #1 \rangle}

\newcommand{\al}{\alpha}
\newcommand{\be}{\beta}

\newcommand{\ora}{\overrightarrow}
\newcommand{\mc}{\mathcal}

\newtheorem{lemma}{Lemma}

\begin{document}

\title{On Possible Indicators of Negative Selection in Germinal Centers}
\author{Bertrand Ottino-Loffler}
\affiliation{Center for Studies in Physics and Biology , Rockefeller University, New York,
New York 10065, USA}
\author{Gabriel D. Victora}
\affiliation{Laboratory of Lymphocyte Dynamics, Rockefeller University,  New York,
New York 10065, USA}
\date{\today}

%<500 words
\begin{abstract}
A central feature of vertebrate immune response is affinity maturation, wherein antibody-producing B cells undergo evolutionary selection in microanatomical structures called germinal centers, which form in secondary lymphoid organs upon antigen exposure. While it has been shown that the median B cell affinity dependably increases over the course of maturation, the exact logic behind this evolution remains vague. Three potential selection methods include encouraging the reproduction of high affinity cells (``birth/positive selection''), encouraging cell death in low affinity cells (``death/negative selection''), and adjusting the mutation rate based on cell affinity (``mutational selection''). While all three forms of selection would lead to a net increase in affinity, different selection methods may lead to distinct statistical dynamics.  We present a tractable model of selection, and analyze proposed signatures of negative selection.  Given the simplicity of the model, such signatures should be stronger here than in real systems.  However, we find a number of intuitively appealing metrics -- such as preferential ancestry ratios, terminal node counts, and mutation count skewness -- are all ill-suited for detecting selection method. 
\end{abstract}

%%%%%%%%%%%%%%5
% We present a model where skew would be most prominent if it were a signal of negative selection, and yet we find no such  signal.  
%%%%%%%%%%%%%%%

\maketitle

%%%%%%%%%%%%%

%Intro
%> Germinal Center as Evolutionary Process
%> Most models of evolution are birth - death
%> Moran Model distinguishes between birth and death
%
%Intro to Model
%> switch to new eta defintion 
%> two-level model makes sense if there is single strong mutation
%> Statistical model to Deterministic Equations for h 
%> Provide h 
%>figure for growth rate
%>rBrD grid 
%
%Reproductive Asym
%> Introduce terminal leaf metric
%> Introduce preferential ancestry metric
%> Show Symmetry in rB rD
%
%Mutation 
%> Skew Hypothesis
%> ODEs for mH mL
%> Show linearization doesn't work
%> Switch to normalized eqs
%> Show V is linear
%> Show S is linear, so Skew -> 0

% O(10^3) cells in GC

%%%%%%%%%%%%%%%%%%%%%%%%%%%%%%%%%%%%%%%%%%%%%%%%%%%%%%%%
%%%%%%%%%%%%%%%%%%%%%%%%%%%%%%%%%%%%%%%%%%%%%%
\section{Introduction}
%%%%%%%%%%%%%%%%%%%%%%%%%%%%%%%%%%%%%%%%%%%%%%

In mathematical biology, the term ``evolution'' has two common associations.  The first is the natural selection of living organisms over time~\cite{darwin2016origin} -- the second is the algorithmic optimization of some fitness variable~\cite{vrugt2007improved, back1993overview, bartz2014evolutionary, yu2010introduction, whitley1996evaluating}.  At times, these two meanings are conflated.  In both, a system contains species of diverse fitnesses, with fitter genotypes persisting longer in time, and mutation generating novel genotypes. In the case of the computational algorithm, ``fitness'' is a well defined attribute which directs the dynamics to a concrete objective. However, real living systems have no universal definition of fitness, so it is hard to frame it as an optimization process. However, there is one place where natural and algorithmic evolution coexist -- affinity maturation. 

A vertebrate’s body is exposed to an endless onslaught of pathogens, to which it responds by producing a large variety of tailored antibodies that bind to and ultimately neutralize these threats. Antibodies are initially generated by a genetic reshuffling process known as V(D)J recombination~\cite{tonegawa1983somatic}.  While the efficacy of these initial antibodies is poor, during infection the body starts generating higher and higher quality antibodies~\cite{eisen1964variations, ada1987clonal}, thanks to process known as {\it affinity maturation}.  Affinity maturation is a process which occurs in germinal centers (GCs), microanatomical structures that form within secondary lymphoid organs (e.g. lymph nodes, spleen) upon exposure to antigen.  Here, a diverse population of B cells (each producing their own antibody) is subjected to evolutionary pressure and high levels of mutation~\cite{mckean1984generation}.  The objective is to find an antibody with a high binding affinity to the target antigen\footnotetext[1]{There is also interest in the topic of broadly neutralizing antibodies~\cite{sok2013effects, doria2015strategies, ferretti2023universal}, but here we will focus on narrow affinity maximization.}~\cite{Note1, victora2012germinal, berek1991maturation}. 

In essence, the immune system is running an evolutionary optimization algorithm, with fitness corresponding to antigen-antibody binding affinity. But what is unknown is the {\it actual} algorithm. Historically, some GC models used birth-limited selection (also known as positive selection), where high fitness cells have accelerated division rates, whereas others used death-limited selection (also known as negative selection), where high fitness cells have diminished death rates~\cite{amitai2017population, gitlin2014clonal, liu1989mechanism, meyer2021molecular}. Presently, there is empirical evidence for birth-selection, but a demand for evidence of death selection~\cite{victora2012germinal}. 

A naive way to model fitness is to just treat it as the difference of birth and death rates, making it a 1D variable~\cite{murray1993mathematical}. However, there are many contexts in which it is important to make fitness multidimensional. For example, in the world of network Moran models, there is a split between birth-selective and death-selective models~\cite{kaveh2015duality, yagoobi2023categorizing}. Despite what the 1D perspective would imply, these seemingly equivalent Moran models can have different outcomes.  This is starkly true in fractional takeover times (the time for a single strain to take over x\% of the population), where swapping the selection method can cause drastic distributional changes~\cite{ottino2017evolutionary}. 

Unfortunately, results for network Moran model takeover times are not  amiable to GCs. Not only do GCs lack a convenient graph structure, but Moran takeover times are usually calculated in a low-mutation limit, whereas GCs feature Somatic Hyper-Mutation (SHM). Thanks to SHM, a B cell's antibody-encoding gene experience on average one mutation per $10^3$ base pairs per division, a million times the baseline rate~\cite{victora2012germinal, mckean1984generation, young2021unique}.

Interestingly, there is some evidence that this mutation rate is not constant. So-called ``clonal bursts'' may occur in germinal centers, where a single B cell divides over and over again with apparently low rates of mutation~\cite{tas2016visualizing, victora2012germinal}. This suggests the possibility of there being {\it mutational selection}. All things being equal, having a lower mutation rate leads to longer-lasting strains throughout the generations.  Having a differential mutation rate between high and low affinity B cells would introduce a third form of selection, independent of birth or death selection. 

In this paper, we will build an analytically tractable model for selection that incorporates birth, death, and mutational selection as independent parameters, rendering a multidimensional fitness. For the sake of having our results be generalizable to other, non-GC evolutionary systems, we will avoid incorporating detailed germinal center mechanics, such as cyclic reentry, interactions with follicular dendritic cells or helper T cells, and receptor-antigen molecular dynamics~\cite{oprea2000dynamics, yaari2015mutation, victora2012identification, bannard2013germinal, victora2012germinal}.  

By building such a simple model, we avoid messy confounding factors, so all signals of negative selection should appear with maximum fidelity.   Despite this, we find many seemingly valid signatures of negative selection to produce no meaningful signal whatsoever.  We will start by introducing the model, as well as its steady state.  Then, we examine patterns in how different genotypes emerge.  And finally, we examine the statistics of the mutation count distribution.  We conclude that while many of the static quantities we evaluate fail to act as a signal of negative selection, many dynamical quantities we identify could potentially work.

%%%%%%%%%%%%%%%%%%%%%%%%%%%%%%%%%%%%%%%%%%%%%%
%%%%%%%%%%%%%%%%%%%%%%%%%%%%%%%%%%%%%%%%%%%%%%
%%%%%%%%%%%%%%%%%%%%%%%%%%%%%%%%%%%%%%%%%%%%%%
\section{Model of Generic Selection}

We will develop a reduced model of $n$ cells evolving in a well-mixed environment. For the sake of simplicity, we will assume there to be only two distinct affinity phenotypes, high affinity (H) and low affinity (L). This is not an unprecedented restriction~\cite{oprea1997somatic}, especially since there are cases where a single mutation can increase affinity by a factor of ten~\cite{allen1988antibody}, essentially rendering the rest of the genome into a high-dimensional neutral space.

We use a discrete time formulation, where exactly one event (a division or a death) occurs every timestep. We say that the germinal center has a carrying capacity of $N$, and the event is a birth with probability proportional to $n$, and a death with probability proportional to $n^2/N$ (so the population is stable at $ n = N$, with logistic-style growth~\cite{verhulst1838notice, cramer2002origins2}.) The odds that a cell dies during a death step is inversely proportional to its {\it death fitness}, which is 1 for low affinity cells and $r_D \geq 1$ for high affinity cells. Similarly, the odds that a cell divides during a birth step is proportional to its {\it birth fitness}, which is 1 for low affinity cells and $r_B \geq 1$ for high affinity cells. That is, $r_B$ controls the level of positive selection, and $r_D$ controls negative selection. We will take $r_B$ and $r_D$ to be finite in the main body of the text, and cover the infinite cases in appendix B in the supplemental information.  

Since we may have mutational selection, the mutation rates for one affinity line may be larger than the other. On division, H cells will have a mutation rate of $\rho_H$, with a $\eta_H$ chance of the mutant being low affinity and $1-\eta_H$ chance of remaining high affinity (thus being a neutral mutation). Similarly, L cells have a mutation rate of $\rho_L$, with a $\eta_L$ chance of the mutant being high affinity, and a $1-\eta_L$ chance of remaining low affinity. It is often more useful to use the net transfer rates, so
\begin{center}
\begin{tabular}{ c c }
$\alpha_H = \rho_H ( 1- \eta_H)$,& $\beta_H = \rho_H \eta_H$, \\
$\alpha_L = \rho_L ( 1- \eta_L)$, & $\beta_L = \rho_L \eta_L$. 
\end{tabular}
\end{center}
$\alpha$ is the rate of neutral mutations, that is, H$\to$H and L$\to$ L mutations.  Meanwhile $\beta$ is the rate of mutating into a different affinity H$\to$L and L$\to$H. Note $\alpha_H + \beta_H = \rho_H$ and $\alpha_L + \beta_L = \rho_L$. 

On a technical note, cellular division is typically symmetric, so one would expect both daughter cells to have equal odds of being a mutant $\rho_X$. However, somatic hypermutation in the germinal center is powered by the action of AID on a single strand of DNA at a time.  So it is possible for one daughter cell to have a mutation rate that is a million times greater than its sister cell~\cite{petersen2002aid, muramatsu1999specific}. With that in mind, and for the sake of notational cleanness, we will only allow one daughter cell of the two to be a potential mutant, with the other being a copy of the mother cell. In appendix A, we cover the generic case of arbitrary mutation rates for both daughters.

%%%%%%%%%%%%%%%%%%%%%%%%%%%%%%%%%%%%%%%%%%%%%%
%%%%%%%%%%%%%%%%%%%%%%%%%%%%%%%%%%%%%%%%%%%%%%
%%%%%%%%%%%%%%%%%%%%%%%%%%%%%%%%%%%%%%%%%%%%%%

\section{Net Selection}

The most immediate quantity to calculate is the overall level of selection -- that is, what is the final percentage of high affinity cells, $h=$ number of H cells/$N$?

Per timestep, let $P_B$ be the probability of a birth/division, and $P_D$ the probability of death. So,
\begin{alignat*}{1}
P_B &= \frac{1}{1+n/N}, \\
P_D &= \frac{n/N}{1+n/N}.
\end{alignat*}

To find the probability of a certain strain dividing, we let $\ell = 1 - h$ and use
\begin{alignat*}{1}
Z_B =~& \frac{P_B}{r_B h + \ell}, \\
P_{BH} =~& r_B h Z_B, \\
P_{BL} =~& \ell Z_B,
\end{alignat*}
where $Z_B$ is just a normalization factor, and the individual probabilities are proportional to the strain's population and birth fitness. 

Similarly, to find the probability of a certain strain having a death, we use
\begin{alignat*}{1}
Z_D =~& \frac{P_D}{h/r_D + \ell}, \\
P_{DH} =~& (h/r_D) Z_D, \\
P_{DL} =~& \ell Z_D,
\end{alignat*}
where $Z_D$ is a normalization factor, and the individual probabilities are proportional to the strain's population and inversely proportional to its death fitness. 

We will now use these probabilities to estimate the average change in $h= \#$ H cells/$n$ over time. First, the probability that the H population goes down by one is simply the probability that an H cell dies, so $P_{-1} = P_{DH}$. To find the probability of a new H cell appearing, we need to consider three sources: an H cell divides (w/o mutation), an H cell divides (with an H$\to$ H mutation), and an L cell divides (with an L $\to$ H mutation). Putting these together, we get:
\begin{alignat*}{1}
P_{+1} =~& P_{BH}(1-\rho_H) + P_{BH} \alpha_H + P_{BL} \beta_L \\
=~& r_B h Z_B (1-\beta_H) + \ell Z_B \beta_L. 
\end{alignat*}

The average population of high affinity cells changes by $P_{+1} - P_{-1}$ every time step, so for large $N$ we can approximate by 
\begin{equation}\label{eq_hdot}
\partial_t h = r_B h Z_B (1-\beta_H) + \ell Z_B \beta_L - (h/r_D) Z_D.
\end{equation}

We want to find the steady-state value of $h$, so let's assume we already hit $n \equiv N$, and therefore $P_D = P_B = 1/2$. So we want to solve
\begin{alignat*}{1}
0 = r_B \frac{1-\beta_H}{h(r_B - 1) +1} + \frac{\beta_L(1-h)}{h(r_b + 1} - \frac{h/r_D}{h(1/r_D -1) + 1},
\end{alignat*}
or equivalently,
\begin{alignat*}{1}
0 =~& \left( -r_B \beta_H/r_D + r_B \beta_H - r_B - \beta_L/r_D + \beta_L + 1/r_D\right) h^2 \\
& + \left( r_B - r_B \beta_H + \beta_L/r_D - 2 \beta_L - 1/r_D \right)h \\
&+ \beta_L.
\end{alignat*}

For reasons which will be obvious later, let's assume $h = 1/(1+ r_B g)$. Substituting in and rearranging, we get 
\begin{alignat*}{1}
0 =~& \left( -r_B \beta_H/r_D + r_B \beta_H - r_B - \beta_L/r_D + \beta_L + 1/r_D\right) g^2 \\
& + \left( r_B - r_B \beta_H + \beta_L/r_D - 2 \beta_L - 1/r_D \right) g \\
&+ \beta_L.
\end{alignat*}
Solving for $g$ (and taking the $g \geq 0$ solution) gives 
\begin{alignat*}{1}
g =  \frac{1}{2 \beta_L r_B r_D} \left(1 - \beta_L + r_B r_D (\beta_H - 1) + \sqrt{\hat g(r_B r_D)} \right),
\end{alignat*}
where 
\begin{alignat*}{1}
\hat g(x) =~& 1 + \beta_{L}^2 + x^2 (1+ \beta_{H}^2) + 2 x \beta_H (1-x) \\
& + 2 \beta_L (-1+x) - 2x (1 - \beta_H\beta_L). 
\end{alignat*}
Therefore, 
\begin{equation} \label{eq_h}
h(r_B, r_D) = \frac{1}{1 + r_B g(r_Br_D)},
\end{equation}
with the special case at $r_B = r_D = 1$ being
\begin{equation} \label{eq_h1}
h(1, 1) = \frac{\beta_L}{\beta_H + \beta_L}.
\end{equation}

\begin{figure}[t]
\centering
\includegraphics[width = 0.5\textwidth]{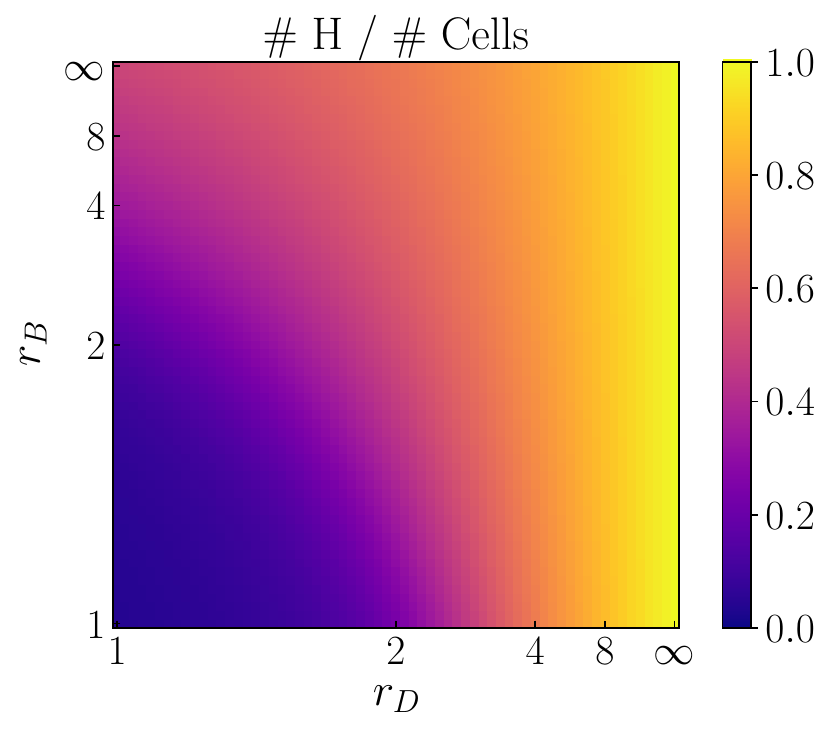}
\caption{Heat map of $h$ via equation~\eqref{eq_h}. Here, $\alpha_H = 0.196$, $\alpha_L = 0.679$, $\beta_H = 0.504$, and $\beta_L = 0.021$. Axes are plotted according to $1 - 1/r_X$ to span from one to infinity. }
\label{fig:hgrid}
\end{figure}

Notably, while $g$ is symmetric in $r_B$ and $r_D$, the net level of selection $h$ is asymmetric in birth and death selectivity, as seen in figure~\ref{fig:hgrid}.  Consider the case where $r_B \gg 1$ and $r_D = 1$, so high affinity cells are dividing at a much higher rate than low affinity cells. However, because the transition rate $\beta_H$ is nonzero, many of those new cells will be low affinity as well, and will stick around for a while because the negative selection pressure is nonexistent. In the limit of large $r_B$, then 
\begin{alignat*}{1}
\lim_{r_B \to \infty} h(r_B, r_D)  = \left(1 + \frac{\beta_H}{r_D (1-\beta_H)} \right)^{-1} < 1. 
\end{alignat*}
This means that there is an effective {\it ceiling of selection} for positive selection, wherein you will always have a mixture of high and low affinity, not matter how selective the system is. Meanwhile, with negative selection, 
\begin{alignat*}{1}
\lim_{r_D \to \infty} h(r_B, r_D) = 1.
\end{alignat*}
So negative selection has no ceiling whatsoever. 

In particular, if a real-world evolutionary system is observed to have 
\begin{equation*}
h_{\mbox{measured}} > h(\infty, 1) =  1 - \beta_H,
\end{equation*}
then said system {\bf must have negative selection}, since the force of positive selection alone is insufficient to attain that level of pressure.

Returning to the specific case of germinal centers, it is well established that a diversity of affinity levels persists. In experiments, even after three weeks, cells can be observed with a {\it lower} affinity than the initial seed population~\cite{araki2023replaying}.  With this logic, if negative selection is present in germinal centers it should not be extremely strong, though this may also be an artifact of the gap between division and selection being long enough for us to observe cells slated for death.

%%%%%%%%%%%%%%%%%%%
%\begin{figure}[h]
%\centering
%\includegraphics[width = 0.5\textwidth]{h_plot.pdf}
%\caption{Plot of $h$ converging to the predicted value from equation~\eqref{eq_h}. Here, $N = 2.5e5$, $r_B = 10$, $r_D= 5$, $\alpha_H = 0.196$, $\alpha_L = 0.679$, $\beta_H = 0.504$, and $\beta_L = 0.021$. Simulation was ran over $T = 1.5e7$ steps, for a total of 60 generations. }
%\label{fig:hvt}
%\end{figure}
%%%%%%%%%%%%%%%%%%%

%%%%%%%%%%%%%%%%%%
%%%%%%%%%%%%%%%%%

%%%%%%%%%%%%%%%%%%%
%\begin{figure}[h]
%\centering
%\includegraphics[width = 0.5\textwidth]{hfinal_contourSMD.pdf}
%\caption{Contour plot of $h$ via equation~\eqref{eq_h}. Here, $\alpha_H = 0.196$, $\alpha_L = 0.679$, $\beta_H = 0.504$, and $\beta_L = 0.021$. Axes are plotted according to $1 - 1/r_X$ to span from one to infinity. }
%\label{fig:hgrid_contour}
%\end{figure}
%%%%%%%%%%%%%%%%%%

%%%%%%%%%%%%%%%%%%%%%%%%%%%%%%%%%%%%%%%%%%%%%%
%%%%%%%%%%%%%%%%%%%%%%%%%%%%%%%%%%%%%%%%%%%%%%
%%%%%%%%%%%%%%%%%%%%%%%%%%%%%%%%%%%%%%%%%%%%%%

\section{The Ancestry Hypothesis}

While this model may only two affinity levels, we can still have a diversity of genotypic strains via neutral mutations (e.g., high $\to$ high affinity mutation). In principle, we can keep track of where these strains come from, therefore giving each strain an ancestor strain. We assume that the genotype space is large compared to the population of cells, so every new mutation produces a novel strain. That is to say, if strain A has a mutation during division, it will produce strain B, and if strain A has another mutation later on, that second mutation will be distinct from strain B. 

Looking at such phylogenetic trees, one might hypothesize that there may be a hidden signature of negative selection. One candidate signature is {\it preferential ancestry}. That is, we count how many high affinity cells have a high affinity strain as their ancestor, denoted by $F_H$. Similarly, we let $F_L$ be the fraction of low affinity cells with a high affinity strain as an ancestor. 

There are three distinct sources of high affinity cells. Per timestep, we attain an average of $P_{BH} (1 - \rho_H)$ high affinity cells from nonmutating high affinity divisions, $P_{BH} \alpha_H$ from mutating high affinity divisions, and $P_{BL} \beta_L$ from low to high affinity divisions. Similarly, there are two sources of H cells with H ancestors: on average we get $F_H P_{BH} (1-\rho_H)$ such cells from non-mutating divisions, and $P_{BH} \alpha_H$ from H cells mutating into other H cells. 

Taking a steady-state, this means
\begin{alignat*}{1}
F_H = \frac{F_H P_{BH} (1-\rho_H) + P_{BH} \alpha_H}{P_{BH} (1 - \rho_H) + P_{BH} \alpha_H + P_{BL} \beta_L}.
\end{alignat*}
Keeping in mind $P_{BL}/P_{BH} = (1/h-1)/r_B = g(r_Br_D)$, then we have 
\begin{equation}\label{eq_fH}
F_H = \frac{\alpha_H}{\alpha_H + \beta_L g(r_Br_D)}.
\end{equation}
By a similar accounting, we find that
\begin{equation}\label{eq_fL}
F_L = \frac{\beta_H}{\beta_H + \alpha_L g(r_Br_D)}.
\end{equation}

\begin{figure}[h]
\centering
\includegraphics[width = 0.5\textwidth]{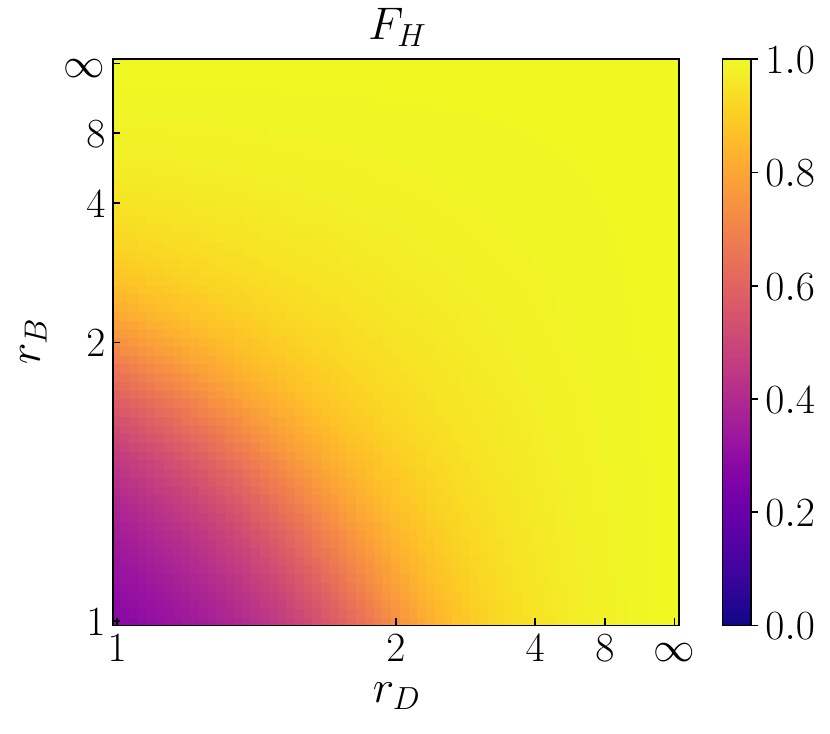}
\caption{Heat map of $F_H$ via equation~\eqref{eq_fH}. Here, $\alpha_H = 0.196$, $\alpha_L = 0.679$, $\beta_H = 0.504$, and $\beta_L = 0.021$. Axes are plotted according to $1 - 1/r_X$ to span from one to infinity. }
\label{fig:fHgrid}
\end{figure}

\begin{figure}[h]
\centering
\includegraphics[width = 0.5\textwidth]{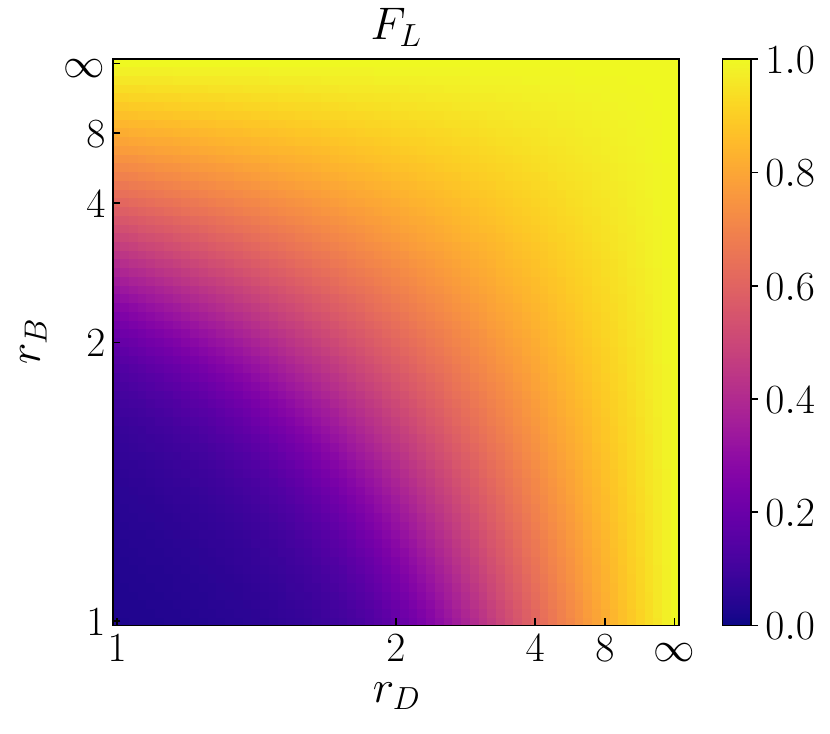}
\caption{Heat map of $F_L$ via equation~\eqref{eq_fL}. Here, $\alpha_H = 0.196$, $\alpha_L = 0.679$, $\beta_H = 0.504$, and $\beta_L = 0.021$. Axes are plotted according to $1 - 1/r_X$ to span from one to infinity. }
\label{fig:fLgrid}
\end{figure}

Notably, both $F_H$ and $F_L$ only depend on positive and negative selection through their product $r_Br_D$. As seen by figures~\ref{fig:fHgrid} and~\ref{fig:fLgrid}, these preferential ancestry ratios are perfectly symmetric in these two fitness parameters. Therefore, in this model, such metrics are ill suited for identifying the presence or absence of negative selection, since it makes no distinction between $r_B$ and $r_D$.

%%%%%%%%%%%%%%%%%%
%%%%%%%%%%%%%%%%%%

%%%%%%%%%%%%%%%%%%
%%%%%%%%%%%%%%%%%%

%
%%%%%%%%%%%%%%%%%%%
%\begin{figure}[h]
%\centering
%\includegraphics[width = 0.5\textwidth]{fH_plot.pdf}
%\caption{Plot of $F_H$ converging to the predicted value from equation~\eqref{eq_fH}. Here, $N = 2.5e5$, $r_B = 10$, $r_D= 5$, $\alpha_H = 0.196$, $\alpha_L = 0.679$, $\beta_H = 0.504$, and $\beta_L = 0.021$. Simulation was ran over $T = 1.5e7$ steps, for a total of 60 generations. }
%\label{fig:fHt}
%\end{figure}
%%%%%%%%%%%%%%%%%%%
%
%
%%%%%%%%%%%%%%%%%%%
%\begin{figure}[h]
%\centering
%\includegraphics[width = 0.5\textwidth]{fL_plot.pdf}
%\caption{Plot of $F_L$ converging to the predicted value from equation~\eqref{eq_fL}. Here, $N = 2.5e5$, $r_B = 10$, $r_D= 5$, $\alpha_H = 0.196$, $\alpha_L = 0.679$, $\beta_H = 0.504$, and $\beta_L = 0.021$. Simulation was ran over $T = 1.5e7$ steps, for a total of 60 generations. }
%\label{fig:fLt}
%\end{figure}
%%%%%%%%%%%%%%%%%%%

\subsection*{Terminal Nodes}

One related metric worth discussing here is the concept of ``terminal nodes.'' As the system of cells continues to divide and reproduce, strains will produce novel mutant descendants, creating a phylogenetic tree of ancestries. At any point of time, this tree will have terminal nodes, which are strains with no extant descendants. It can be conjectured that if most low affinity cells are in terminal nodes, then there ought to be negative selection at work.

\begin{figure}[t]
\centering
\includegraphics[width = 0.5\textwidth]{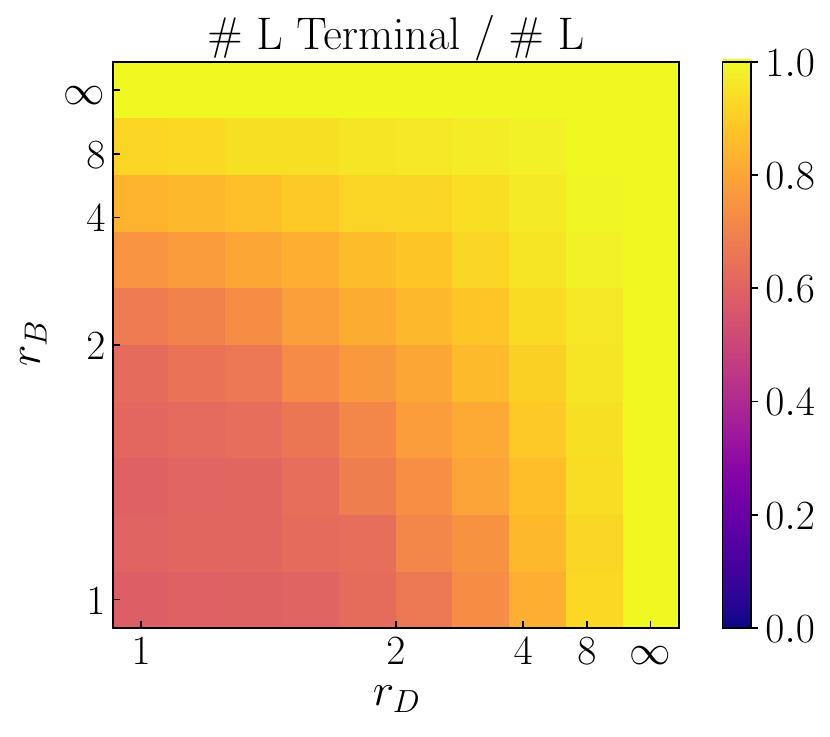}
\caption{Plot of the final fraction of L cells which are in terminal nodes. Here, $N = 5e3$, $\alpha_H = 0.196$, $\alpha_L = 0.679$, $\beta_H = 0.504$, and $\beta_L = 0.021$. Axes are plotted according to $1 - 1/r_X$ to span from one to infinity. Simulation was ran over $T = 3e5$ steps, for a total of 60 generations. We numerically take $0/0 = 1$.  }
\label{fig:terminal_L}
\end{figure}

A proper analytic exploration of this metric is somewhat out of scope of this paper, but we can generate examples numerically. In figure~\ref{fig:terminal_L}, we show the number of L cells in terminal nodes divided by the total number of L cells. Notice that this figure appears symmetric in $r_B$ and $r_D$. Intuitively, this makes sense: if L cells only appear as terminal nodes, then either they die before they divide (high negative selection), or they don't get a chance to divide in the first place (high positive selection). 

Similarly, the total fraction of cells which are in terminal nodes shows poor signal in $r_D$ in figure~\ref{fig:terminal_L}. While it might be tempting to tease out a trend, it should be noted in real-world datasets, phylogenetic trees have to be attained via statistical reconstruction. The simulations here create the tree with perfect knowledge, whereas in practice there will be ambiguity over cell terminality, which would likely drown out whatever slight signal may be present. 

In addition, most other ratios involving number of terminal cells can be derived from the prior two metrics, $f_H$, $F_L$, and the total selection level $h$. While some terminal node metrics may have signals of negative selection, they would simply be from the signal present in the much easier to measure $h$. Moreover, pilot simulations for counting number of genotypes (instead of number of cells) suggest the same results. 

As such, it is hard to recommend any terminal node metric as a measure of negative selection.

%%%%%%%%%%%%%%%%%%
%%%%%%%%%%%%%%%%%%

%%%%%%%%%%%%%%%%%%
\begin{figure}[t]
\centering
\includegraphics[width = 0.5\textwidth]{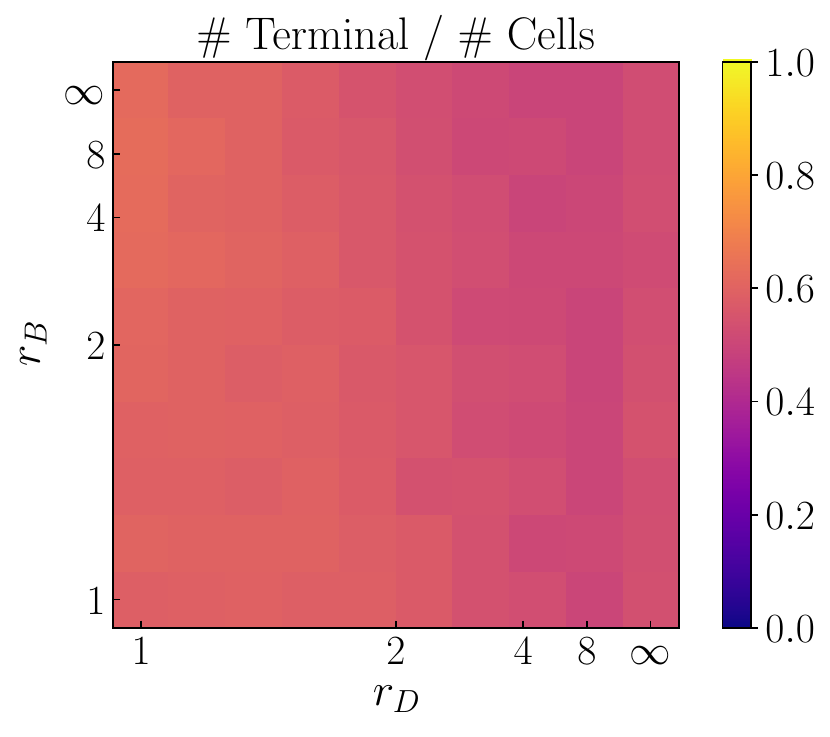}
\caption{Plot of the final fraction of cells which are in terminal nodes. Here, $N = 5e3$, $\alpha_H = 0.196$, $\alpha_L = 0.679$, $\beta_H = 0.504$, and $\beta_L = 0.021$. Axes are plotted according to $1 - 1/r_X$ to span from one to infinity. Simulation was ran over $T = 3e5$ steps, for a total of 60 generations. }
\label{fig:terminal_n}
\end{figure}
%%%%%%%%%%%%%%%%%%

%%%%%%%%%%%%%%%%%%%%%%%%%%%%%%%%%%%%%%%%%%%%%%
%%%%%%%%%%%%%%%%%%%%%%%%%%%%%%%%%%%%%%%%%%%%%%
%%%%%%%%%%%%%%%%%%%%%%%%%%%%%%%%%%%%%%%%%%%%%%

\section{Mutation Count Distribution}\label{sec:mutation}

Even though this model has only two phenotypes (low and high affinity), because of the large neutral space, the number of potential genotypes is extremely high.  As before, we will assert that no two mutation events will ever be identical. That is, it is unlikely two separate mutation events will lead to the exact same base pair sequence.  If we initialize the germinal center with a population of genetically identical cells, we should be able to count how many mutations are accumulated in each cell line. 

If a species grows at a rate $r$ and has a mutation rate $\rho$, then you'd expect the population to accrue mutations at an average rate of $r \rho$. In our model, we have two separate growth rates and mutation rates, $r_B$ \& $\rho_H$ and $1$ \& $\rho_L$, but the actual average accumulation rate is somewhat more complicated due to the L$\to$ H and H $\to$ L mutations coupling the two mutation distributions. 

More to the point, dynamically measuring a population's mutational distribution is not always a good option. For many biological systems, doing a genotypic survey can be expensive and/or destructive (e.g., needing to destroy a germinal center to analyze its B cells), so measuring dynamical properties is usually unappealing. 

\begin{figure}[t]
\centering
\includegraphics[width = 0.5\textwidth]{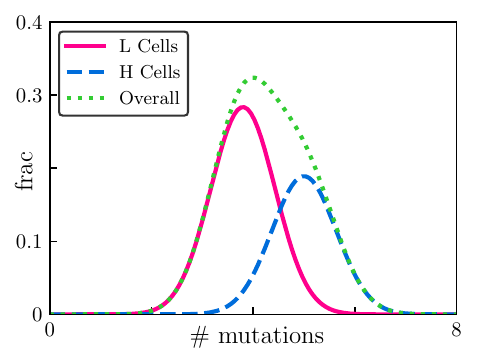}
\caption{Schematic showing the intuition behind the skew hypothesis. Here, different levels of mutational activity causes a split between the high fitness and low fitness B cells, producing skewness in the overall distribution.}
\label{fig:Two_Normals}
\end{figure}

However, if the distribution of mutation counts is different between low and high affinity cells, then intuitively the overall mutation count distribution ought to be asymmetric. That is, we can hypothesize that the mutation count distribution ought to have some level of {\it skew} over long times, as in figure~\ref{fig:Two_Normals}.  Moreover, a back of the envelope calculations suggests that the skew would approach a constant value over time. Skew is calculated via the central moments $C_a$, and a naive examination would suggest $C_a \propto t^a$. Therefore, the skew would look like 
\begin{equation*}
\mbox{Skew} = \frac{C_3}{(C_2)^{3/2}} \approx \frac{t^3}{ (t^{2})^{3/2}} = \mbox{constant}.
\end{equation*}
Because this is a prediction of the long-term shape of the distribution, and not its dynamics, it can be measured and approximated by a single snapshot genotypic survey. The experimental advantage of taking one measurement over a longitudinal study need not be elaborated upon. Moreover, this final skew value should be a function of the different fitness parameters. Therefore, one can hypothesize that measuring mutational skew may be a good indicator of negative selection. 

What follows is an explanation as to why this hypothesis is false.

%%%%%%%%%%%%%%%%%%
%%%%%%%%%%%%%%%%%%

To investigate the mutation count distribution, let's assume that the overall affinity of the population has hit a steady state. Next, let us define $h_m$ to be the number of high affinity cells with $m$ mutations, divided by the total number of cells $n$. Similarly, we let $\ell_m$ be the fraction of cells which are low affinity and have $m$ mutations, and $n_m = h_m + \ell_m$. Note that $\sum_m h_m = h$, and $\sum_m \ell_m = \ell = 1-h$. 

To construct the dynamics for $h_m$, we do some basic accounting. Sources for H cells with $m$ mutations are H and L cells with $m-1$ mutations, as well as H cells with $m$ mutations. In the single-mutant dynamics, the only sink is the natural death rate. Therefore, the dynamics are given by 
\begin{alignat*}{1}
\partial_t h_m =~& \left(1-\rho_H \right) \left(h_m r_B Z_B \right) + \alpha_H \left( h_{m-1} r_B Z_B\right) \\
&+ \beta_L \left(\ell_{m-1} Z_B \right) - (h_m/r_D) Z_D \\
\partial_t \ell_m =~& \left( 1 - \rho_L\right)\left(\ell_m Z_B \right) + \alpha_L \left(\ell_{m-1} Z_B\right) \\
&+ \beta_H \left( h_{m-1} r_B Z_B\right) - \ell_m Z_D. 
\end{alignat*}

To get the central moments $C$, we first need the regular moments.  To find the moments of the mutational distributions, we define the moments $\mc{H}_k := \sum_m m^{k} h_m$ and $\mc{L}_k:= \sum_m m^{k} \ell_m$ for the H and L populations respectively. Taking $\partial_t$ of both sides, we get
\begin{alignat}{4}
\partial_t \mc{H}_k =& \left[ \left(1-\beta_H \right) r_B Z_B - \frac{Z_D}{r_D}\right]\mc{H}_k   + \beta_L Z_B \mc{L}_k  \notag \\
&+\sum_{w=0}^{k-1} {k \choose w} \left(  r_B \alpha_H Z_B  \mc{H}_w + \beta_L Z_B \mc{L}_w \right)  \label{mHdot} \\
\partial_t \mc{L}_k =& \left[ \left(1-\beta_L \right) Z_B -Z_D \right] \mc{L}_k + r_B \beta_H Z_B \mc{H}_k \notag \\
&+ \sum_{w=0}^{k-1} {k \choose w}  (r_B \beta_H Z_B \mc{H}_w + \alpha_L Z_B \mc{L}_w), \label{mLdot} 
\end{alignat}
where we handled $h_{m-1}$ and $\ell_{m-1}$ by changing the sum index and using the binomial theorem, and taking $h_{-1} = \ell_{-1} = 0$.

While unpleasant to look at, this is fundamentally a linear system of equations. Moreover, since each $k$'th moment only depends on moments $k$ to $0$, this gives it a block lower-triangular structure, which should be amenable to analysis.  In principle, a closed-form solution involving matrix exponentials should be possible.  However, in appendix C, we show that this approach is both numerically and theoretically fraught, due to the dynamical matrix becoming poorly conditioned over time.  Therefore, a slightly different approach is needed.

To make analysis easier, we will instead use the following change of variables, and assume at least one of $r_B$ and $r_D$ are larger than 1:
\begin{alignat}{1}
\mathcal{M}_k &:=  \mc{H}_k+ \mc{L}_k, \label{eq_Mdef_main} \\
\mathcal{S}_k &:= (r_B Z_B \beta_H/\ell)  \mc{H}_k + (Z_B \beta_L/h)  \mc{L}_k. \label{eq_Sdef_main}
\end{alignat}
$\mathcal{M}_k$ represents the $k$'th moment of the mutation count distribution for the overall population, including both H and L cells.  Meanwhile, the definition of $\mathcal{S}_k$ was just chosen to make the dynamics cleaner, as given by
\begin{alignat}{1}
\partial_t \mathcal{M}_k =~& -\tau \mathcal{M}_k + \mathcal{S}_k \notag \\
& + M_M \sum_{w = 0}^{k-1} {k \choose w} \mc{M}_w + M_S \sum_{w = 0}^{k-1} {k \choose w} \mc{S}_w, \label{eq_Mdot} \\
\partial_t \mathcal{S}_k =~& S_M \sum_{w = 0}^{k-1} {k \choose w} \mc{M}_w + S_S \sum_{w = 0}^{k-1} {k \choose w} \mc{S}_w \label{eq_Sdot}.  
\end{alignat} 
The prefactors are given by   
\begin{alignat*}{1}
M_M =~& Z_B \frac{\rho_L \beta_H - r_B \rho_H \beta_L g}{\beta_H - \beta_L g}, \\
M_S =~& \frac{r_B \rho_H - \rho_L}{r_B \beta_H/\ell - \beta_L/h},
\end{alignat*}
and 
\begin{alignat*}{1}
S_M =~& \frac{r_B Z_{B}^2 \beta_H \beta_L} {r_B \beta_H h - \beta_L \ell} \left( \beta_H/g - r_B \beta_L g + \alpha_L - r_B \alpha_H \right), \\
S_S =~& Z_B \frac{r_B \beta_H \alpha_H + \beta_H \beta_L(r_B g - 1) - \alpha_L \beta_L g}{\beta_H - g \beta_L}.
\end{alignat*}
$\tau$ is given by
\begin{alignat*}{1}
\tau = Z_B \left( \frac{\beta_H}{g} + \beta_L r_B g \right). 
\end{alignat*}
Notice that $\tau > 0$ always, so it functions as a proper timescale for the system.

%%%%%%%%%%%%%%%%%%
\begin{figure}[t]
\centering
\includegraphics[width = 0.5\textwidth]{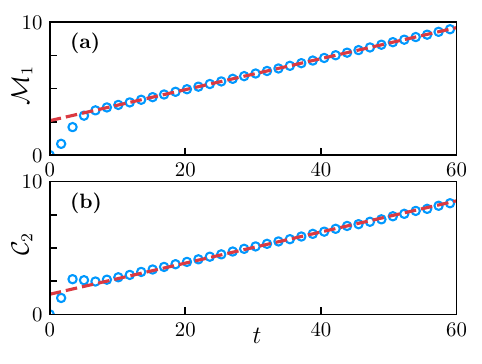}
\caption{Plot of the mutation count's (a) average $\mc{M}_1$ and (b) variance $C_2$. he circles are  values  attained from simulation, using $N = 2.5e5$, $r_B = 10$, and $r_D = 5$, with $\alpha_H = 0.196$, $\alpha_L = 0.679$, $\beta_H = 0.504$, and $\beta_L = 0.021$. The dashed line indicates the slope as predicted by~\eqref{eq_mu} in (a), and by~\eqref{eq_Cta-1} in (b). }
\label{fig:M1C2}
\end{figure}
%%%%%%%%%%%%%%%%%%

%%%%%%%%%%%%%%%%%%
\begin{figure}[h]
\centering
\includegraphics[width = 0.5\textwidth]{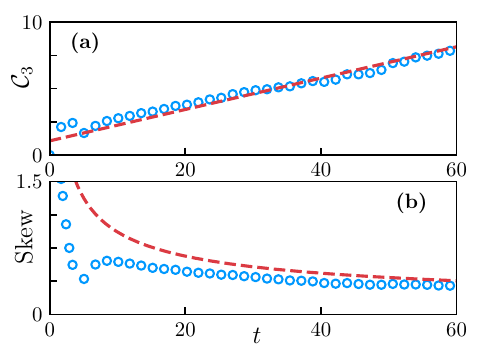}
\caption{Plot of the mutation count's (a) third central moment $C_3$ and (b) skew. The circles are  values attained from simulation, using $N = 2.5e5$, $r_B = 10$, and $r_D = 5$, with $\alpha_H = 0.196$, $\alpha_L = 0.679$, $\beta_H = 0.504$, and $\beta_L = 0.021$. The dashed line in (a) indicates the predicted slope of $C_3$ (via appendix D), and in (b) a $1/\sqrt{t}$ curve to illustrate the gradual decay, as predicted by equation~\eqref{eq_skew_limit}.}
\label{fig:C3skew}
\end{figure}
%%%%%%%%%%%%%%%%%%

Also note that $\mathcal{S}_k$ only depends on lower order terms, meaning that it is possible to get exact solutions by iteratively solving.  Both $\mc{S}_k$ and $\mc{M}_k$ will be polynomials in time $t$ with a maximum degree $k$, plus some other terms that decay exponentially with rate $\tau$.  

Looking at the definitions of $\mc{M}$ and $\mc{S}$ in~\eqref{eq_Mdef_main} and~\eqref{eq_Sdef_main}, it is easy to see $\mc{M}_0 = 1$ and $\mc{S}_0 = \tau$.  By solving the differential equations for $k=1$, we get 
\begin{alignat*}{1}
\mc{S}_1(t) =~& \tau \mu t + \theta^{S}_1, \\
\mc{M}_1(t) =~& \mu t + \left( M_M + \tau M_S + \theta^{S}_1\right)/\tau \notag \\
& + \left(\theta^{M}_1 - ( M_M + \tau M_S + \theta^{S}_1)/\tau \right) e^{-\tau t},
\end{alignat*}
where the $\theta$ terms are just initial conditions, and $\mu := S_M/\tau + S_S$. That is to say, the growth rate of the average number of mutations over time is $\mu$, which can be rewritten as
\begin{equation}\label{eq_mu}
\mu = r_B Z_B \frac{ \alpha_H \beta_H /g + \alpha_L \beta_L g + 2 \beta_H \beta_L}{\beta_H/g + r_B \beta_L g}. 
\end{equation}

Omitting the details of the recursion (which can be found in appendix D), we get that the leading order behavior for the $k$'th moment is given by 
\begin{alignat}{1}
\mc{M}_k[t^k] =~& \mu^k. \label{eq_qMa_final}
\end{alignat}
As expected, the $k$'th moment grows like $t^k$, and the leading rate follows a simple form. The expression for the first correction (which is needed to calculate the distribution's skew) is somewhat more unwelcoming, with
\begin{alignat}{1}\label{eq_qMa-1_final}
\mc{M}_k&[t^{k-1}] = \frac{k}{\tau} \mu^{k-1} \theta^{S}_1 + k \mu^{k-2} \left[\frac{(k-1) }{2} \mu \right. \notag \\
 & \left. + \left( \frac{S_M}{\tau} (k-1) + \mu \right)\left( \frac{M_M}{\tau} + M_S - \frac{\mu}{\tau} \right) \right].
\end{alignat}

In order to calculate the mutation distribution's skew and variance, we will need the central moments.  The $a$'th central moment is defined by
\begin{alignat*}{1}
C_a =~& \Exp{(m - \mc{M}_1)^a}\\
=~& \sum_{k=0}^a {a \choose k} \mc{M}_k {\mc{M}_1}^{a-k} (-1)^{a-k}.
\end{alignat*}
Just as with the regular moments, these should be polynomials of maxium degree $a$ and decaying exponential corrections.

We don't need the full solution for the central moment, just the leading growth term.  But by some careful manipulation (also found in appendix D), we find that the supposed ``leading'' term is always zero,
\begin{equation}\label{eq_Cta}
C_a[t^a] = 0.
\end{equation}

When we try the next leading order term, we similarly find for $a \not= 2$ that 
\begin{equation}\label{eq_Cta-1_general}
C_a[t^{a-1}] = 0.
\end{equation}
However, for the special case of $a=2$, we get
\begin{equation}\label{eq_Cta-1}
C_2[t^{1}] = 
\mu + \frac{2 S_M}{\tau} \left(\frac{M_M}{\tau} + M_S - \frac{\mu}{\tau} \right).
\end{equation}
While $C_2$ has a $t^1$ term, we find $C_3$ has no $t^2$ term, $C_4$ has no $t^3$ term, and so on.  

In particular, we now know that the variance (given by $C_2$) and the third central moment ($C_3$) must grow linearly in time. This is corroborated by simulation in figures~\ref{fig:M1C2}a,\ref{fig:M1C2}b, and \ref{fig:C3skew}a. 

With this knowledge, we can finally estimate the skew of the mutant distribution.  Plugging into the definition of skew, we find 
\begin{equation}\label{eq_skew_limit}
\mbox{Skew} = \frac{C_3}{{C_2}^{3/2}} = \frac{\mathbb{O}(t)}{\mathbb{O}(t^{3/2})} \to 0.
\end{equation}
This is supported by simulation results in~\ref{fig:C3skew}b, where the skew indeed decays as $1/\sqrt{t}$.  Therefore, there will be no long-term skew in the mutation count histogram, regardless of the choice of fitnesses.  And so, the hypothesis outlined at the start of the section is false.

%%%%%%%%%%%%%%%%%%
\begin{figure}[t]
\centering
\includegraphics[width = 0.5\textwidth]{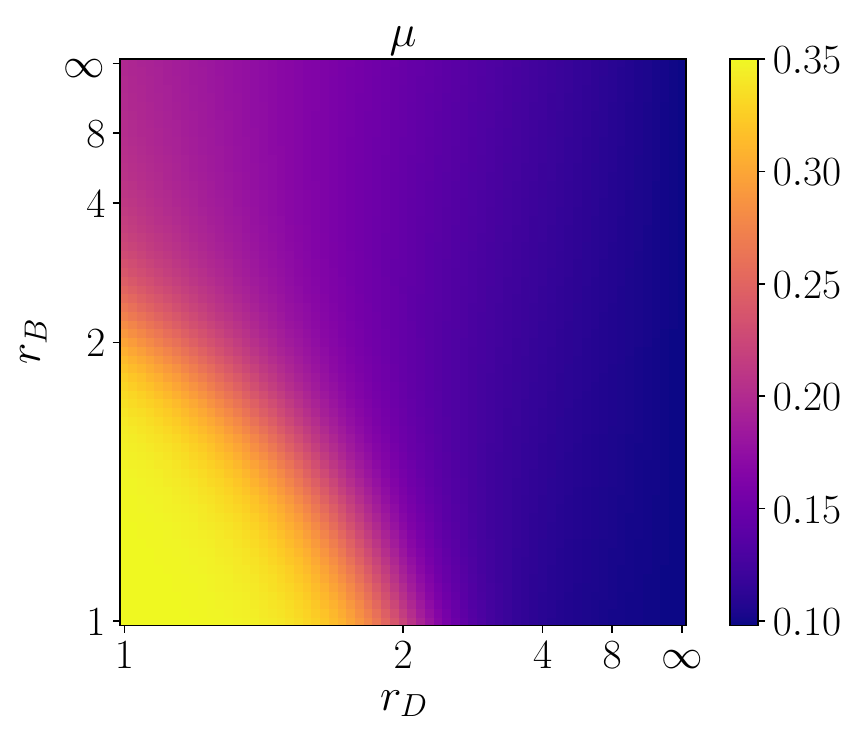}
\caption{Heat diagram of the growth rate $\mu$ of the average mutation count $\mc{M}_1$, via equation~\eqref{eq_mu}. Here, $\alpha_H = 0.196$, $\alpha_L = 0.679$, $\beta_H = 0.504$, and $\beta_L = 0.021$. Axes are plotted according to $1 - 1/r_X$ to span from one to infinity.}
\label{fig:mu}
\end{figure}
%%%%%%%%%%%%%%%%%%

%%%%%%%%%%%%%%%%%%
\begin{figure}[h]
\centering
\includegraphics[width = 0.5\textwidth]{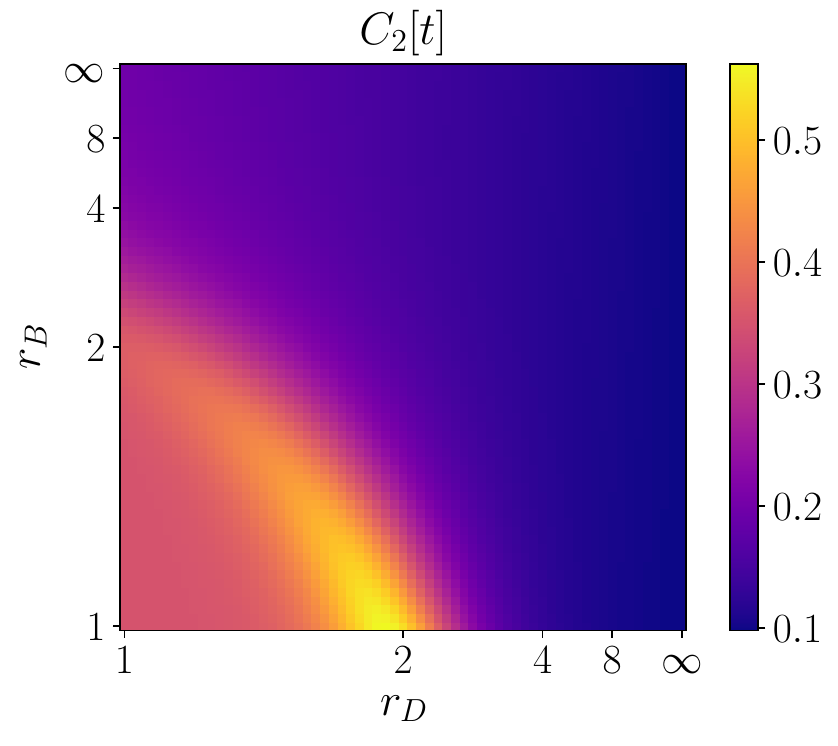}
\caption{Heat diagram of the growth rate $C_2[t]$ of the mutation count's variance, via evaluating equation~\eqref{eq_Cta-1} at $a=2$. Here, $\alpha_H = 0.196$, $\alpha_L = 0.679$, $\beta_H = 0.504$, and $\beta_L = 0.021$. Axes are plotted according to $1 - 1/r_X$ to span from one to infinity.}
\label{fig:q1_C2}
\end{figure}
%%%%%%%%%%%%%%%%%%

%%%%%%%%%%%%%%%%%%%%%%%%%%%%%%%%%%%%%%%%%%
%%%%%%%%%%%%%%%%%%%%%%%%%%%%%%%%%%%%%%%%%%
%%%%%%%%%%%%%%%%%%%%%%%%%%%%%%%%%%%%%%%%%%

\section{Discussion}

%
%
%
%%%%%%%%%%%%%%%%%%%
%\begin{figure}[t]
%\centering
%\includegraphics[width = 0.5\textwidth]{q1_C3_versusrBrDSMD.pdf}
%\caption{Heat diagram of the growth rate of the mutation count's third central moment $c_3$. Here, $\alpha_H = 0.196$, $\alpha_L = 0.679$, $\beta_H = 0.504$, and $\beta_L = 0.021$. Axes are plotted according to $1 - 1/r_X$ to span from one to infinity.}
%\label{fig:q1_C3}
%\end{figure}
%%%%%%%%%%%%%%%%%%%

Uncovering the mechanisms behind an evolutionary process is key -- not only for advancing understanding and recreating natural systems~\cite{sepkoski2016replaying, gould1989wonderful}, but also for improving our own optimizing algorithms~\cite{vrugt2007improved, sharma2022suppressors, tkadlec2020limits}, and providing avenues for better medical interventions~\cite{young2021unique}.   

Based on our work, certain proposed metrics for measuring such mechanisms are ill suited for the task. Preferential ancestry ratios and mutational count asymmetry are particularly poor at the task, since in stable limits, all signature of mechanism vanishes. 

The net selectivity, $h$, is asymmetric in positive and negative fitness. However, it remains problematic for practical use. Outside of extreme cases, interpreting values of $r_B$ and $r_D$ out of a single $h$ would require detailed knowledge of mutation rates. However, since an evolutionary system may employ time-variable mutational adaptation (e.g., a cell line mutating rapidly while exposed to stress), rates measured in vitro may differ in vivo. Many other metrics we could propose here would have similar problems, or are simply redundant with $h$.  For example, if we were to take the average number of mutations in L cells and divide by the average among H cells, the ratio would converge to $r_B g$.  While this is certainly asymmetric, it gives no new information compared to the more accessible metric $h$.

This is not to say that no signature of mechanism exist -- it is just that snapshot-style metrics have difficulty capturing dynamical differences. For example, the growth rate of average mutation count expresses mild asymmetry in figure~\ref{fig:mu}, which only grows more pronounced when looking at the variance in figure~\ref{fig:q1_C2}. However, genotypic surveys, especially for microscopic cells, tend to be highly invasive, so the experimental burden of measuring any of these quantities may be too great to recommend. 

However, perfect symmetry is rare in the real world. In an actual examination, one might expect that a quantity such as $F_H$ would have some asymmetry, or that the mutational skew would take on some long-term value, simply by the vagaries of nature. While the germinal center may have many black boxes, there are many in silica and in vitro experiments one can do which have highly controlled evolutionary mechanisms, some of which may contradict such perfect results. However, the model presented here is highly generic, especially considering the results from appendix A. If such a discrepancy were to be acknowledged, then it would cast doubt on many classes of mathematical evolutionary models.

One final complication specific to GCs may lie in the details of cyclic reentry. The essence of this scheme is that B cells move back and forth between two zones within the GC -- a ``dark zone'' where division takes place, and a ``light zone'' where B cells are evaluated for affinity~\cite{oprea2000dynamics, yaari2015mutation, victora2010germinal, bannard2013germinal, victora2012germinal}. Many evolutionary models treat reproduction and fitness measurement as simultaneous. Here, there is ostensibly a period of ``fitness blindness,'' where cells live and divide unaware of their own affinity/fitness, introducing noise in the fitness landscape~\cite{trubenova2019surfing, merrell1994adaptive}. Moreover, depending on how the mutation-proffering AID is used, the actual mutation rates of cell lines may wildly fluctuate, leading to hypothetical scenarios where the number of mutations is unproportional to the number of divisions.  If that is the case, we may need to use a multistage, seasonal growth model~\cite{swartz2022seascape}.

%diagram of flows into/out of h_m?

%%%%%%%%%%%%%%%%%%%%%%%%%%%%%%%%%%%%%%%%%%%%%%
\section*{Acknowledgements}

This research was supported by the Center for Studies in Physics and Biology at Rockefeller University (B.O.-L.). We thank Arup Chakraborty, Kevin O'Keffee, Daniel Abrams, and Juhee Pae for the helpful comments.  

\section*{Author Contributions}
Bertrand Ottino-Loffler provided the main text, mathematical analysis, and numerics.  Gabriel D. Victora provided the biological background and motivation.

\bibliography{GerminalCenters_Bib}{}

%%%%%%%%%%%%%%%%%%%%%%%%%%%%%%%%%%%%%%%%%%%%%
%%%%%%%%%%%%%%%%%%%%%%%%%%%%%%%%%%%%%%%%%%%%%
%%%%%%%%%%%%%%%%%%%%%%%%%%%%%%%%%%%%%%%%%%%%%

\pagebreak
\widetext
\begin{center}
\title{Supplemental Information: On Possible Indicators of Negative Selection in Germinal Centers}
\end{center}
%%%%%%%%%% Merge with supplemental materials %%%%%%%%%%
%%%%%%%%%% Prefix a "S" to all equations, figures, tables and reset the counter 
\setcounter{equation}{0}
\setcounter{figure}{0}
\setcounter{table}{0}
\setcounter{page}{1}
\makeatletter
\renewcommand{\theequation}{S\arabic{equation}}
\renewcommand{\thefigure}{S\arabic{figure}}
\renewcommand{\bibnumfmt}[1]{[S#1]}
\renewcommand{\citenumfont}[1]{S#1}

\appendix

%%%%%%%%%%%%%%%%%%%%%%%%%%%%%%%%%%%%%
\section{Asymmetric Mutation Rates}\label{sec:VMD}
In this section, we will show that having non-identical mutation rates between the daughter cells produces the same results. 

In particular, if an H cell divides, its first daughter has a $\alpha_{H1}$ chance of mutating into another H cell and a $\beta_{H1}$ chance of becoming a L cell, and the second daughter has $\alpha_{H2}$ and $\beta_{H2}$. Similarly, the individual daughters of an L cell have an L$\to$L transition rate of $\alpha_{L1}$ and $\alpha_{L2}$, and an L$\to$H transition rate of $\beta_{L1}$ and $\beta_{L2}$ respectively. We define the total transition rates as $\bar \beta_X = \beta_{X1}+\beta_{X2}$, $\bar \alpha_X = \alpha_{X1}+\alpha_{X2}$, $ \rho_{Xn} = \beta_{Xn}+ \alpha_{Xn}$, and $\bar \rho_{X} = \bar \beta_{X}+ \bar \alpha_{X}$. To avoid trivial fixed points, we require $\bar \beta_H$ and $\bar \beta_L$ to be nonzero. 

Note that while we require $\bar \beta_H$ and $\bar \beta_L$ to be nonzero, nothing prevents, say, $\beta_{H1} = \alpha_{H1}=0$. Here, one daughter would always be indistinguishable from the parent, which recreates the case from the main paper and some models of sexual reproduction. Similarly, if we want a more typical asexual reproduction model, setting the two daughters to have identical rates will do that for you.  

Also, while instances of extreme mutation can be interesting, we shall also require $\bar \be_H$ and $\bar \be_L$ to be less than 1 to avoid some pathological regimes. For example, if $\bar \be_H > 1$, then $r_B \to \infty$ would unintuitively cause the high fitness population to go extinct, since $h$ decreases per division. 

On a division event, the total population of high fitness (H) cells can change by +2, +1, 0, -1, depending on the identity of the parent and the offspring. For example, if a low fitness (L) cell divides, and the first daughter mutates into a H cell and the second daughter is a nonmutated L cell, this would cause +1 new H cells. We notate this event as ($L \to H^*L$), where the $*$'s identify the daughters as mutants, and the order tracks the identity of the daughters. Using this notation, we can write down all the H-changing events as 
\begin{alignat*}{1}
P_{+2} =~& \left( L \to H^*H^* \right), \\
P_{+1} =~& \left(H \to HH \right) + \left(H \to H^*H \right) + \left(H \to HH^* \right) + \left(H \to H^*H^* \right) \\
& + \left(L \to H^*L \right) + \left(H \to LH^* \right) + \left(L \to H^*L^* \right) + \left(H \to L^*H^* \right), \\
P_{-1} =~& \left(H \to L^*L^*\right) + \left(H \to \mbox{Death} \right).
\end{alignat*}
Given that $h$ is the fraction of the population which is H cells and $\ell$ is the fraction which is L cells, we can write the above events as probabilities to get
\begin{alignat*}{1}
 \partial_t h =~& 2P_{+2} + P_{+1} - P_{-1}\\
=~& r_B h Z_B \left[(1-\rho_{H1})(1-\rho_{H2}) + \alpha_{H1}(1-\rho_{H2}) + (1-\rho_{H1})\alpha_{H2} + \alpha_{H1}\al_{H2} - \be_{H1}\be_{H2}\right] \\
&+ \ell Z_B \left[ \be_{L1}\be_{L2} + \be_{L1}(1-\rho_{L2}) + (1-\rho_{L1}) \be_{L2} + \al_{L1}\be_{L2} + \be_{L1} \al_{L2} \right] - (h/r_D) Z_D,
\end{alignat*}
where $Z_B$ and $Z_D$ are defined as before. This simplifies into 
\begin{equation*}
\partial_t h = r_B h Z_B (1 - \bar \be_H) + \ell Z_B \bar \be_L - (h/r_D) Z_D.
\end{equation*}
This is identical to the $\partial_t h$ equation in the single mutation case (section III in the main body), so all the results from that section would apply with the substitution $\be_H \to \bar \be_H$ and and the like. In particular, 
\begin{equation*}
h = \frac{1}{1+ r_B g(r_Br_D)}.
\end{equation*} 

If we are careful about accounting cells in the same manner as about, we retrieve familiar ancestry ratios $f_H$ and $f_L$
\begin{alignat*}{1}
f_H =~& \frac{\bar \al_H}{\bar \al_H + \bar \be_L g(r_Br_D) }, \\
f_L =~& \frac{\bar \be_H}{\bar \be_H + \bar \al_L g(r_Br_D) }. 
\end{alignat*}
We also get familiar equations for the number of mutations within strains, with
\begin{equation*}
\begin{cases}
\partial_t h_m = \left(1-\bar \rho_H \right) \left(h_m r_B Z_B \right) + \bar \alpha_H \left( h_{m-1} r_B Z_B\right) + \bar \beta_L \left(\ell_{m-1} Z_B \right) - \frac{h_m}{r_D} Z_D, \\
\partial_t \ell_m = \left( 1 - \bar \rho_L\right)\left(\ell_m Z_B \right) + \bar \alpha_L \left(\ell_{m-1} Z_B\right) + \bar \beta_H \left( h_{m-1} r_B Z_B\right) - \ell_m Z_D. 
\end{cases}
\end{equation*}
Hence, all results from the main body also apply to the general case of asymmetrically mutating daughters.

%%%%%%%%%%%%%%%%%%%%%%%%%%%%%%%%
\section{Infinite fitness edge cases}\label{sec:Infinite}

Throughout this paper we considered only finite fitnesses, but on various heat plots, we include the case of $r_B = \infty$ and $r_D = \infty$. Here we will discuss the approach for each. Moreover, for the sake of completeness, we will allow both daughters to be mutants, as in appendix~\ref{sec:VMD}. 

\subsection{Limit of large positive selection.}

In the case of infinite birth fitness, only high fitness cells divide. So $P_{BH} = P_B$ and $P_{BL} = 0$ so long as $h > 0$. In such as case,
\begin{alignat*}{1}
P_{+2} =~& 0,\\
P_{+1} =~& \left(H \to HH \right) + \left(H \to H^*H \right) + \left(H \to HH^* \right) + \left(H \to H^*H^* \right) \\
=~& P_{BH} \left( 1 - \bar \beta_H + \beta_{H1} \be_{H2}\right), \\
P_{-1} =~&\left(H \to L^*L^* \right) + \left(H \to \mbox{Death} \right) \\
=~& P_{BH} \be_{H1}\be_{H2} + P_{DH}.
\end{alignat*}
This means the average growth rate of high-fitness cells becomes 
\begin{equation*}
\partial_t h = \frac{1}{2} \left(1-\bar \be_H\right) - \frac{h}{r_D} Z_D. 
\end{equation*}
The steady state can be solved from here to get
\begin{equation*}
h = \frac{r_D ( 1-\bar \be_H) }{r_D ( 1-\bar \be_H) + \bar \be_H }.
\end{equation*}

Similarly, the equations for the mutant populations also get simplified, with 
\begin{alignat*}{1}
\partial_t h_m =~& \frac{h_m}{2h} (1 - \bar \rho_H), \\
\partial_t \ell_m =~& \frac{h_{m-1}}{2h} \bar\be_H - Z_D \ell_m. 
\end{alignat*}
While simpler, the structure is similar to that in section V in the main body, so we can just take the results from there and take the limit of $r_B \to \infty$ to quickly get the results. Notably, we have
\begin{alignat*}{1}
\lim_{r_B \to \infty} Z_B =~& 0 \\
\lim_{r_B \to \infty} r_B g =~& \frac{\bar \beta_H}{r_D (1- \bar\be_H)}, \\
\lim_{r_B \to \infty} Z_B/g =~& \frac{r_B}{2h} \frac{1-\bar \beta_H}{\bar \beta_H},\\
\lim_{r_B \to \infty} r_B Z_B =~& \frac{1}{2h}. 
\end{alignat*}
As applied to the building blocks of equations for $ \partial_t \mc{M}_k$ and $ \partial_t \mc{S}_k$ (from section V in the main body), we have 
\begin{alignat*}{1}
&\tau = r_D \frac{1-\beta_H}{2h}, \\
&M_M = 0, \\ 
&M_S = \ell \frac{\bar \rho_H}{\bar \beta_H}, \\
&S_M = 0,\\
&S_S = \frac{\bar \alpha_H}{2h}. 
\end{alignat*}
Although some terms go to zero, this does not disrupt the rest of the analysis, leading to the same conclusion about the skew approaching zero. 

\subsection{Limit of large negative selection.}
Although the limit of large positive selection was routine, a little more care must be taken with large negative selection. For example, trying to naively take the $r_D \to \infty$ limit to the results of the main body's section V leads to the timescale $\tau$ diverging. 

Because of the high level of negative selection, so long as any L cells exist, no H cells will die. In fact, an attempt to write an equation for $h$ yields 
\begin{equation*}
\partial_t h = \ell Z_B \bar \be_L + h (1- \bar\be_H) Z_B. 
\end{equation*}
Notably, this is strictly positive so long as $\bar\be_H < 1$. So, absent finite-size errors, $h \to 1$ is assured. 

Because there is no low-fitness population worth talking about, we have that the H population is the full population, so $\mc{H}_k = \mc{M}_k$. In particular, we have
\begin{equation*}
\partial_t h_M = \frac{\bar \alpha_H}{2} (h_{m-1} - h_m),
\end{equation*} 
where the $1/2$ term comes from the fact that at steady state, $P_B = P_D = 1/2$. We therefore have 
\begin{equation*}
\partial_t \mc{M}_k = \frac{\alpha_H}{2} \sum_{w=0}^{k-1} {k \choose w} \mc{M}_w.
\end{equation*} 
This is recursively solvable, recalling that $\mc{M}_0 = 1$. Letting $\gamma := \bar\al_H/2 $, the first three orders are given by 
\begin{alignat*}{1}
\mc{M}_1 =~& \gamma t + \theta_{1}^M , \\
\mc{M}_2 =~& \gamma^2 t^2 + (2 \gamma \theta_{1}^M + \gamma) t + \theta_{2}^M ,\\
\mc{M}_3 =~& \gamma^3 t^3 + 3 \gamma^2 ( \theta_{1}^M + 1) t^2 + \gamma (3 \theta_{2}^M + 3 \theta_{1}^M +1)t + \theta_{3}^M.\\
\end{alignat*}
Therefore, the variance is given by 
\begin{alignat*}{1}
C_2 =~& \mc{M}_2 - (\mc{M}_1)^2\\
=~& \gamma t + \theta_{2}^M - \left(\theta_{1}^M\right)^2,
\end{alignat*}
and the third central moment is given by 
\begin{alignat*}{1}
C_3 =~& \mc{M}_3 - 3 \mc{M}_2\mc{M}_1 + 2 (\mc{M}_1)^3 \\
=~& \gamma t + \theta^{M}_3 - 3 \theta^{M}_2 \theta^{M}_1 + 2 ( \theta^{M}_1)^3. 
\end{alignat*}
That is to say, the mean, variance, and third central moment all grow with at the exact same rate of $\gamma = \bar\al_H/2$. Therefore, the skew goes as 
\begin{equation*}
\mbox{Skew} = \frac{C_3}{(C_2)^{3/2}} = \frac{\gamma t + \mbox{const.}}{(\gamma t + \mbox{const.})^{3/2}} \to 0. 
\end{equation*}

So, just as in the main case, we have no long-term skew appearing.

%%%%%%%%%%%%%%%%%%%%%%%%%%%%%%%%
\section{The Matrix Exponential Solution}\label{sec:matrix_exp}

In the main body, we got equations for the moments of L and H populations 
\begin{alignat}{4}
\partial_t \mc{H}_k &= \left[ \left(1-\beta_H \right) r_B Z_B - \frac{Z_D}{r_D}\right] \mc{H}_k &&+ \beta_L Z_B \mc{L}_k &&+ r_B \alpha_H Z_B \sum_{w=0}^{k-1} {k \choose w} \mc{H}_w &&+ \beta_L Z_B \sum_{w=0}^{k-1} {k \choose w} \mc{L}_w \label{mHdot} \\
\partial_t \mc{L}_k &= \left[ \left(1-\beta_L \right) Z_B -Z_D \right] \mc{L}_k &&+ r_B \beta_H Z_B \mc{H}_k &&+ r_B \beta_H Z_B \sum_{w=0}^{k-1} {k \choose w} \mc{H}_w &&+ \alpha_L Z_B \sum_{w=0}^{k-1} {k \choose w} \mc{L}_w. \label{mLdot} 
\end{alignat}

Let us vectorize the system in equations~\eqref{mHdot} and~\eqref{mLdot} by letting $\overrightarrow m = (\mc{H}_1, \mc{L}_1, \mc{H}_2,\mc{L}_2,\mc{H}_3, \mc{L}_3)^T,$ where we truncate to $k=3$, since our goal is to find the skew. Therefore, the dynamics are given by
\begin{equation*}
\partial_t \ora m = J \ora{m} + \ora{v},
\end{equation*}
where the odd entries of $\ora{v}$ are $h r_B \alpha_H Z_B + \beta_L Z_B \ell,$ and the even entries are $h r_B \beta_H Z_B + \alpha_L Z_B \ell$. Meanwhile, $J$ is a block matrix of the form:
\begin{equation*}
J=
\begin{pmatrix}
J_D & 0 & 0 \\
2 J_{S} & J_D & 0 \\
3 J_{S} & 3 J_{S} & J_D 
\end{pmatrix},
\end{equation*}
with 
\begin{equation*}
J_D=
\begin{pmatrix}
r_B Z_B (1-\beta_H) - Z_D/r_D & \beta_L Z_B \\
r_B Z_B \beta_H & Z_B (1-\beta_L) - Z_D 
\end{pmatrix},
\end{equation*}
and 
\begin{equation*}
J_S=
\begin{pmatrix}
r_B \alpha_H Z_B & \beta_L Z_B \\
r_B Z_B \beta_H & \alpha_L Z_B
\end{pmatrix}.
\end{equation*}
So long as the determinant of $J$ is nonzero, this means that the moments are, in principle, exactly solvable, via
\begin{equation*}
\ora{m}(t) = e^{J t} (\ora{m}(0) + J^{-1} \ora v) - J^{-1} \ora v.
\end{equation*}

The determinant of $J$ is decided by $J_D$. So calculating $|J_D|$ gives 
\begin{alignat*}{1}
|J_D| = r_B Z_{B}^2(1-\beta_H)(1-\beta_L) - r_B Z_B Z_D (1-\beta_H) - Z_B Z_D (1-\beta_L) /r_D + Z_{D}^2/r_D - r_B \beta_H \beta_L Z_{B}^2.
\end{alignat*}
If we calculate at population steady-state ($n=N$), then we have
\begin{alignat*}{1}
\frac{Z_B}{Z_D} &= 1+ 2 \left(r_{D}^{-1} - r_B\right) h Z_B, \\ 
\frac{Z_D}{Z_B} &= 1+ 2 \left(r_{D}^{-1} - r_B\right) h Z_D. 
\end{alignat*}
Therefore, 
\begin{alignat*}{1}
|J_D|Z_{B}^{-1}Z_{D}^{-1} =~& r_B \frac{Z_B}{Z_D} \left[ (1-\beta_H)(1-\beta_L) - \beta_H \beta_L\right] - r_B(1-\beta_H) - (1-\beta_L)\frac{1}{r_D} + \frac{1}{r_D} \frac{Z_D}{Z_B} \\ 
=~& r_B (1-\beta_H -\beta_L)\left(1 + 2 (r_{D}^{-1}- r_B)h Z_B \right) + \frac{1}{r_D} \left(1 + 2 (r_{D}^{-1} - r_B)h Z_D \right) - r_B(1-\beta_H) - (1-\beta_L) \frac{1}{r_D} \\
=~& r_B (1-\beta_H) - r_B \beta_L + 2 r_B (1-\beta_L -\beta_H) \left(r_{D}^{-1} - r_B\right) h Z_B + \frac{1}{r_D} \\
&+ 2 \left(r_B - r_{D}^{-1}\right) \frac{h}{r_D} Z_D - r_B(1-\beta_H) - (1-\beta_L)\frac{1}{r_D} \\
=~& \left(r_{D}^{-1} - r_B\right) \left( \beta_L + 2 r_B (1-\beta_L -\beta_H) h Z_B - 2 h Z_D/r_D \right) \\
=~& \left(r_{D}^{-1} - r_B\right) \left( 2 r_B h Z_B ( 1- \beta_H) - 2 h Z_D/r_D + 2 \ell Z_B \beta_L - 2 \ell Z_B \beta_L + \beta_L - 2r_B \beta_L h Z_B \right) \\
=~& \left(r_{D}^{-1} - r_B\right) \left( 2 \partial_t h + \beta_L\left[1 - 2 Z_B(r_B h + \ell) \right] \right) \\
=~& \left(r_{D}^{-1} - r_B\right) \left( 2 \partial_t h + \beta_L\left[1 - 1\right] \right), 
\end{alignat*}
and so we get 
\begin{equation*}
|J_D| = 2 \left(r_{D}^{-1} - r_B\right) Z_B Z_D \partial_t h.
\end{equation*}
As the system approaches a phenotypic equilibrium, we have $ \partial_t h \to 0$. So if we want to find long-time dynamics (or worse, stationary state dynamics), then this vectorized formulation is ill suited for numerical and analytical examination.

\section{Recusion for Mutant Moments}\label{sec:mutant_recursion}
In this section, we will produce the recursion equations that define the growth of the moments of the mutant distribution  $\mathcal{M}_k$, and give the leading-order results for the central moments $C_a$.  

Specifically, we have the following system of equations
\begin{alignat}{1}
\partial_t \mathcal{M}_k =~& -\tau \mathcal{M}_k + \mathcal{S}_k + M_M \sum_{w = 0}^{k-1} {k \choose w} \mc{M}_w + M_S \sum_{w = 0}^{k-1} {k \choose w} \mc{S}_w, \label{eq_Mdot} \\
\partial_t \mathcal{S}_k =~& S_M \sum_{w = 0}^{k-1} {k \choose w} \mc{M}_w + S_S \sum_{w = 0}^{k-1} {k \choose w} \mc{S}_w \label{eq_Sdot}, 
\end{alignat} 
where 
\begin{alignat}{1}
\mathcal{M}_k &:= \mc{H}_k+ \mc{L}_k, \label{eq_Mdef} \\
\mathcal{S}_k &:= (r_B Z_B \beta_H/\ell) \mc{H}_k + (Z_B \beta_L/h)  \mc{L}_k, \label{eq_Sdef}
\end{alignat}
for all $k \geq 1$.  

By hypothesis, we assume the solutions take the form of polynomials, with
\begin{equation}\label{eq_Sgeneral}
\mc{S}_k = \sum_{w=0}^k q^{Sk}_w t^w + \sum_{w=0}^{k-2} p^{Sk}_{w} t^w e^{-\tau t},
\end{equation}
and 
\begin{equation}\label{eq_Mgeneral}
\mc{M}_k = \sum_{w=0}^k q^{Mk}_w t^w + \sum_{w=0}^{k-1} p^{Mk}_{w} t^w e^{-\tau t}, 
\end{equation}
where $q^{Mk}_w$, $p^{Mk}_{w}$, $q^{Sk}_w$, and $p^{Sk}_{w}$ are coefficients to be determined, and we take coefficients with index $ w < 0$ to be zero. 

Just as in the main text, the $k = 1$ step is easy to verify. We start with the $\mc{S}_1$ equation, and substitute appropriate values for $\mc{M}_0$ and $\mc{S}_0$ based on their definitions in~\eqref{eq_Mdef} and~\eqref{eq_Sdef}, and get
\begin{alignat*}{1}
\partial_t \mathcal{S}_1 =~& S_M \mc{M}_0 + S_S \mc{S}_0 \\
=~& S_M (h +\ell) + S_S ( r_B Z_B \beta_H h/\ell + Z_B \beta_L \ell/h) \\
=~& S_M + S_S \tau.
\end{alignat*}
For future notational convenience, we define 
\begin{equation}\label{eq_mu_first}
\mu := S_M/\tau + S_S.
\end{equation}
Therefore, we have $\mc{S}_1(t) = \tau \mu t + \theta^{S}_1$, where we use $\theta^{S}_k := \mc{S}_k(t=0)$ and $\theta^{M}_k := \mc{M}_k(t=0)$. 

Similarly, we have 
\begin{alignat*}{1}
\partial_t \mathcal{M}_1 =~& -\tau \mc{M}_1 + \mc{S}_1 + M_M \mc{M}_0 + M_S \mc{S}_0 \\
=~& -\tau \mc{M}_1 + q^{S1}_1 t + (M_M + \tau M_S) + q^{S1}_0.
\end{alignat*}
Integrating gives us 
\begin{equation*}
\mc{M}_1 = \mu t + \left( M_M + \tau M_S + \theta^{S}_1\right)/\tau + \left(\theta^{M}_1 - ( M_M + \tau M_S + \theta^{S}_1)/\tau \right) e^{-\tau t},
\end{equation*}
so to leading order, the mean mutation count grows at a rate $\mu := S_M/\tau + S_S$.

With a base established, we can create the inductive step to establish a recursion relation for the various coefficients. By subbing equations~\eqref{eq_Mgeneral} and~\eqref{eq_Sgeneral} into~\eqref{eq_Sdot}, we get
%\begin{alignat*}{1}
%\partial_t \mc{S}_{a+1} =~& S_M \sum_{k=0}^a {a+1 \choose k} \mc{M}_k + S_S \sum_{k=0}^a {a+1 \choose k} \mc{S}_k \\
%=~& S_M \sum_{k=0}^a {a+1 \choose k} \sum_{w=0}^k q^{Wk}_w t^w + S_M \sum_{k=0}^a {a+1 \choose k} \sum_{w=0}^{k-2} p^{Wk}_w t^w e^{-\tau t} \\
%& + S_S \sum_{k=0}^a {a+1 \choose k} \sum_{w=0}^k q^{Sk}_w t^w + S_S \sum_{k=0}^a {a+1 \choose k} \sum_{w=0}^{k-2} p^{Sk}_w t^w e^{-\tau t} \\
%\end{alignat*}
\begin{alignat*}{1}
\partial_t \mc{S}_{a+1} = \sum_{k=0}^a d_{k}^{S(a+1)} t^k + \sum_{k=0}^{a-1} c_{k}^{S(a+1)} t^k e^{-\tau t}
\end{alignat*}
where
%\begin{alignat*}{1}
%d_{k}^{S(a+1)} &= \sum_{w = k}^a {a+1 \choose w} \left(S_M q_{k}^{Mw} + S_S q_{k}^{Sw} \right)
%c_{k}^{S(a+1)} &= (a+1) S_M p^{Ma} + S_M \sum_{w=k}^{a-2} {a+1 \choose w+1} p_{k}^{M(w+1)} + S_S \sum_{w=k}^{a-2} {a+1 \choose w+2} p_{k}^{S(w+2}; k \leq a-2 \\
%c_{a-1}^{S(a+1)} &= (a+1) S_M p_{a-1}^{Ma} 
%\end{alignat*}
\begin{alignat}{1}
d_{k}^{S(a+1)} &= \sum_{w = k}^a {a+1 \choose w} \left(S_M q_{k}^{Mw} + S_S q_{k}^{Sw} \right) \label{eq_dSk} \\
c_{k}^{S(a+1)} &= (a+1) S_M p^{Ma} + S_M \sum_{w=k}^{a-2} {a+1 \choose w+1} p_{k}^{M(w+1)} + S_S \sum_{w=k}^{a-2} {a+1 \choose w+2} p_{k}^{S(w+2)} \label{eq_cSk}
\end{alignat}
where we used lemma~\ref{lem_sumorder} in appendix~\ref{sec:Lemmas} to collect terms appropriately, and we take the convention that $\sum_{w=x}^y z_w = 0$ when $ x > y$. 

If we plug these into the expressions in lemma~\ref{lem_ind} in appendix~\ref{sec:Lemmas}, we get 
\begin{alignat*}{1}
\mc{S}_{a+1} = \sum_{k=0}^{a+1} q_{k}^{S(a+1)} t^k + \sum_{k=0}^{a-1} p_{k}^{S(a+1)} t^k e^{-\tau t} 
\end{alignat*}
%\begin{alignat*}{1}
%q_{k}^{S(a+1)} &= \frac{1}{k}\sum_{w=k-1}^a {a+1 \choose w} \left[ S_M q_{k-1}^{Mw} + S_S q_{k-1}^{Sw} \right],
%\end{alignat*}
%with the $k=0$ coefficient given by 
%\begin{alignat*}{1}
%q_{0}^{S(a+1)} =~& \theta_{a+1}^S + \frac{(a-1)!}{\tau^a} (a+1) S_M p^{Ma}_{a-1} \\
%&+ \sum^{a-2}_{w = 0} \frac{w!}{\tau^{w+1}} \left[(a+1) S_M p_{w}^{Ma} + \sum_{r=w}^{a-2} \left\{ S_S {a+1 \choose r+2} p_{w}^{S(r+2)} + S_M { a+1 \choose r+1} p_{w}^{M(r+1)} \right\} \right],
%\end{alignat*}
%and the coefficients for the exponential terms are 
%\begin{alignat*}{1}
%p_{k}^{S(a+1)} =~& -\frac{(a-1)!}{\tau^{a-k}} \frac{(a+1)}{k!} S_M p^{Ma}_{a-1} \\
%& - \sum^{a-2}_{w = k} \frac{w!}{k! \tau^{w-k+1}} \left[(a+1) S_M p_{w}^{Ma} + \sum_{r=w}^{a-2} \left\{ S_S {a+1 \choose r+2} p_{w}^{S(r+2)} + S_M { a+1 \choose r+1} p_{w}^{M(r+1)} \right\} \right].
%\end{alignat*}
with 
\begin{alignat}{1}
q_{k}^{S(a+1)} =~& \frac{1}{k} d_{k-1}^{S(a+1)}, k\geq 1 \label{eq_qSk}\\
q_{0}^{S(a+1)} =~& \theta_{a+1}^S + \sum_{w=0}^{a-1} \frac{w!}{\tau^{w+1}} c_{k}^{S(a+1)} \label{eq_qS0} \\
p_{k}^{S(a+1)} =~& -\sum_{w=k}^{a-1} \frac{w!}{k! \tau^{w+1}} c_{k}^{S(a-1)}. \label{eq_pSk} 
\end{alignat}

By a similar procedure, we have that 
\begin{alignat*}{1}
\partial_t \mc{M}_{a+1} + \tau \mc{M}_a = \sum_{k=0}^{a+1} d_{k}^{M(a+1)} t^k + \sum_{k=0}^{a-1} c_{k}^{M(a+1)} t^k e^{-\tau t}
\end{alignat*}
with 
\begin{alignat}{1}
d_{k}^{M(a+1)} =~& q_{k}^{S(a+1)} + \sum_{w = k}^{a} {a+1 \choose w} \left( M_M q_{k}^{Mw} + M_S q^{Sw}_k \right) \label{eq_dMk} \\ 
c_{k}^{M(a+1)} =~& p_{k}^{S(a+1)} + M_{M} (a+1) p_{k}^{M a} + \sum_{w = k}^{a-2} \left[ {a+1 \choose w+1} p_{k}^{M(w+1)} M_M + {a+1 \choose w+2} p_{k}^{S(w+2)} M_S \right]. \label{eq_cMk}
\end{alignat}
If we plug these into the expressions in lemma~\ref{lem_nonind} in appendix~\ref{sec:Lemmas}, we then get 
\begin{alignat*}{1}
\mc{M}_{a+1} = \sum_{k=0}^{a+1} q_{k}^{M(a+1)} t^k + \sum_{k=0}^{a} p_{k}^{M(a+1)} t^k e^{-\tau t} 
\end{alignat*}
with 
\begin{alignat}{1}
q^{M(a+1)}_k =~& \sum_{w=k}^{a+1} \frac{w!}{k!} \frac{(-1)^{w-k}}{\tau^{w-k+1}} d_{w}^{M(a+1)} \label{eq_qMk} \\
p^{M(a+1)}_k =~& \frac{1}{k} c_{k-1}^{M(a+1)}, k\geq1 \label{eq_pMk} \\
p^{M(a+1)}_0 =~& \theta_{a+1}^{M} + \sum_{w=0}^{a+1} w! \frac{(-1)^{w+1}}{\tau^{w+1}} d_{w}^{M(a+1)}. \label{eq_pM0}
\end{alignat}

If we insert~\eqref{eq_cMk} and~\eqref{eq_dMk} into equation~\eqref{eq_qMk} and evaluate at $k = a+1$, we get 
\begin{equation}\label{eq_qMa+1}
q_{a+1}^{M(a+1)} = \frac{1}{\tau} q_{a+1}^{S(a+1)}.
\end{equation}
Similarly, if we insert~\eqref{eq_cSk} and~\eqref{eq_dSk} into equation~\eqref{eq_qSk} and evaluate at $k = a+1$, we find 
\begin{equation}\label{eq_qSa+1}
q_{a+1}^{S(a+1)} = S_M q_{a}^{Ma} + S_S q_{a}^{Sa}.
\end{equation}
Putting equations~\eqref{eq_qMa+1} and~\eqref{eq_qSa+1} together gives the following closed forms via induction, using the fact that $\mu = S_S + S_M/\tau$:
\begin{alignat}{1}
q_{a}^{Ma} =~& \mu^a, \label{eq_qMa_final} \\
q_{a}^{Sa} =~& \tau \mu^a. \label{eq_qSa_final}
\end{alignat}

If we instead evaluate~\eqref{eq_qSk} and~\eqref{eq_qMk} at $k = a$, we get 
\begin{alignat*}{1}
q_{a}^{S(a+1)} =~& \frac{a+1}{2} \tau \mu^a + \frac{a+1}{a} \left( S_M q_{a-1}^{Ma} + S_S q_{a-1}^{Sa} \right), \\
q_{a}^{M(a+1)} =~& \frac{1}{\tau} q_{a}^{S(a+1)} + \frac{\mu^a}{\tau}(a+1) \left( M_M + M_S \tau - \mu \right). 
\end{alignat*}
Naturally, we can insert one into the other to find
\begin{equation*}
q_{a}^{S(a+1)} = \frac{a+1}{a} \mu q_{a-1}^{Sa} + \mu^{a-1} (a+1) \left( \frac{ \mu \tau}{2} + S_M \left[ M_M/\tau + M_S - \mu/\tau \right] \right).
\end{equation*}
If we then solve this via the lemma~\ref{lem_Iter} in appendix~\ref{sec:Lemmas}, we get for $a \geq 1$
\begin{equation*}
q_{a-1}^{Sa} = a \mu^{a-1} q_{0}^{S1} + a(a-1)\mu^{a-2} \left(
\mu \tau/2 + S_M\left[ M_M/\tau + M_S - \mu/\tau \right] \right). 
\end{equation*}
Therefore, the subleading coefficient for the $a$th moment is
\begin{equation}\label{eq_qMa-1_final}
q_{a-1}^{Ma} = \frac{a}{\tau} \mu^{a-1} q_{0}^{S1} + a \mu^{a-2} \left[\frac{(a-1) \mu}{2} + \left( \frac{S_M}{\tau} (a-1) + \mu \right)\left( \frac{M_M}{\tau} + M_S - \frac{\mu}{\tau} \right) \right].
\end{equation}

To recap, equation~\eqref{eq_qMa_final} is the leading order coefficient of the growth of the $a$th moment of mutation counts, and \eqref{eq_qMa-1_final} is the subleading coefficient. This gives us enough information to calculate the behavior of the central moments, which characterize the shape of the mutation count histogram. 

The $a$th central moment is given by 
\begin{alignat*}{1}
C_a =~& \Exp{(m - \mc{M}_1)^a}\\
=~& \Exp{ \sum_{k=0}^a {a \choose k} m^k {\mc{M}_1}^{a-k} (-1)^{a-k} }\\
=~& \sum_{k=0}^a {a \choose k} \mc{M}_k {\mc{M}_1}^{a-k} (-1)^{a-k}.
\end{alignat*}
We take $a \geq 2$, since the first central moment is definitionally zero. 

If we substitute in the polynomial forms of the $\mc{M}_k$ terms and collect all the exponential terms under $\mc{E}$, we get 
\begin{alignat*}{1}
C_a =~& \sum_{k=0}^a {a \choose k} \left( \sum_{w=0}^k q_{w}^{Mk} t^w \right) \left( q_{0}^{M1} + \mu t\right)^{a-k} (-1)^{a-k} + \mc{E} 
\end{alignat*}
If we collect terms, this takes on the following form:
%\begin{alignat}{1}
%& C_a = \sum_{b=0}^a t^b \sum_{k=0}^a {a \choose k} (-1)^{a-k} \sum_{(w, r) \in I(a, b, k)} f_{a, k}(w,r) \\
%& f_{a, k}(w,r) = {a - k \choose w} \left(q_{0}^{M1} \right)^{a-k-w} q_{r}^{Mk} \mu^w \\
%& I(a, b, k) = \{(w, r) | 0 \leq w \leq a-k, 0 \leq r \leq k, w + r = b \}
%\end{alignat}
\begin{alignat}{1}
& C_a = \sum_{b=0}^a t^b \sum_{k=0}^a {a \choose k} (-1)^{a-k} \sum_{w = \max(0, b-k)}^{\min(b,a-k)} f_{a, k}(w,b-w) + \mc{E} , \label{eq_CaFull} \\
& f_{a, k}(w,r) = {a - k \choose w} \left(q_{0}^{M1} \right)^{a-k-w} q_{r}^{Mk} \mu^w, \notag
\end{alignat}
where the bounds on the sum are from the first argument of $f$ needing to be less than $a-k$, the second has to be less than $k$, and their sum must be $b$. 

Via the general form of the moments $\mc{M}_a$ (Equation~\eqref{eq_Mgeneral}), we know the leading term of the central moments grows at most as $t^a$. So to get the proper coefficient, we evaluate the $b=a$ term to get 
\begin{alignat*}{1}
C_a[t^a] =~& \sum_{k=0}^a {a \choose k} (-1)^{a-k} \sum_{w = \max(0, a-k)}^{\min(a,a-k)} f_{a, k}(w,a-w) \\
=~& \sum_{k=0}^a {a \choose k} (-1)^{a-k} f_{a, k}(a-k,a-(a-k)) \\
=~& \sum_{k=0}^a {a \choose k} (-1)^{a-k} {a - k \choose a-k} \left(q_{0}^{M1} \right)^{a-k-(a-k)} q_{k}^{Mk} \mu^{(a-k)} \\
=~& \sum_{k=0}^a {a \choose k} (-1)^{a-k} q_{k}^{Mk} \mu^{(a-k)} \\
=~& \sum_{k=0}^a {a \choose k} (-1)^{a-k} \mu^k \mu^{(a-k)} \\
=~& \mu^a \sum_{k=0}^a {a \choose k} (-1)^{a-k} 1^k \\
=~& \mu^a (1 - 1)^a.
\end{alignat*}
Note the use of equation~\eqref{eq_qMa_final}. Therefore, we have that for all $a \geq 1$, then 
\begin{equation}\label{eq_Cta}
C_a[t^a] = 0.
\end{equation}
In other words, every central moment grows at most as $t^{a-1}$. 

To find the $a-1$ order coefficient, we pull the $b=a-1$ term from Equation~\eqref{eq_CaFull}. Here, 
\begin{alignat*}{1}
C_a[t^{a-1}] =~& \sum_{k=0}^a {a \choose k} (-1)^{a-k} \sum_{w = \max(0, (a-1)-k)}^{\min(a-1,a-k)} f_{a, k}(w,(a-1)-w) \\
=~& \sum_{k=0}^a {a \choose k} (-1)^{a-k}\left( f_{a, k}(a-k-1, k) + f_{a, k}(a-k,k-1)\right) \\
%=~& {a \choose 0} (-1)^{a} f_{a,0}(a-1, 0) + {a \choose a} (-1)^{a-a} f_{a, a-1}(0, a-1)\\
%& + \sum_{k=1}^{a-1} {a \choose k} (-1)^{a-k}\left( f_{a, k}(a-k-1, k) + f_{a, k}(a-k,k-1)\right) \\
%=~& (-1)^a {a - 0 \choose a-1} \left(q_{0}^{M1} \right)^{a-0-(a-1)} q_{0}^{M0} \mu^{a-1}\\ 
%&+ {a - a \choose 0} \left(q_{0}^{M1} \right)^{a-a-0} q_{a-1}^{Ma} \mu^{0} \\
%&+ \sum_{k=1}^{a-1} {a \choose k} (-1)^{a-k} {a-k \choose k-1} \left(q_{0}^{M1} \right)^{a-k-(a-k-1)} q_{k}^{Mk} \mu^{a-k-1} \\
%&+ \sum_{k=1}^{a-1} {a \choose k} (-1)^{a-k} {a-k \choose a-k} \left(q_{0}^{M1} \right)^{a-k-(a-k)} q_{k-1}^{Mk} \mu^{a-k} \\
=~& \sum_{k=0}^a (a-k) {a \choose k} (-1)^{a-k} q_{0}^{N1} \mu^{a-1} + \sum_{k=1}^a {a \choose k} (-1)^{a-k} q_{k-1}^{Nk} \mu^{a-k},
\end{alignat*}
where we already inserted~\eqref{eq_qMa_final}. If we also insert~\eqref{eq_qMa-1_final}, we find 
\begin{alignat*}{1}
C_a[t^{a-1}] =~& q_{0}^{N1} \mu^{a-1} \sum_{k=0}^a {a \choose k} (-1)^{a-k} (a-k) + q_{0}^{S1} \frac{ \mu^{a-1} }{\tau} \sum_{k=1}^a{ a\choose k} (-1)^{a-k} k \\
& + \mu^{a-2} \left[\frac{\mu}{2} + \frac{S_M}{\tau}\left(\frac{M_M}{\tau} + M_S - \frac{\mu}{\tau} \right) \right] \sum_{k=1}^a {a \choose k} (-1)^{a-k} k (k-1) \\
& + \mu^{a-1} \left[ \frac{M_M}{\tau} + M_S - \frac{\mu}{\tau} \right] \sum_{k=1}^{a} (-1)^{a-k} {a \choose k} k .
\end{alignat*}
While this is unpleasant to look at, all the sums simplify to various Kronecker deltas $\delta_{j,k}$. Keeping in mind $a \geq 2$ and eliminating the appropriate deltas, this simplifies to 
\begin{equation}\label{eq_Cta-1_sup}
C_a[t^{a-1}] = \delta_{a, 2} \left[\mu + \frac{2 S_M}{\tau} \left(\frac{M_M}{\tau} + M_S - \frac{\mu}{\tau} \right) \right].
\end{equation}
This is nonzero when $a = 2$, and zero otherwise.   This returns us to the results in the main text.

%%%%%%%%%%%%%%%%%%%%%%%%%%%%%%%%%%%%
\section{Useful Lemmas}\label{sec:Lemmas}

A number of elementary lemmas appear in our induction proofs, so for ease of reference they have been collected here. 

\begin{lemma} \label{lem_ind}
Given the following equation for $z$, 
\begin{equation*}
\dot z = -\tau z + \sum_{k = 0}^m d_k t^k + \sum_{k = 0}^n c_{k} t^k e^{-\tau t},
\end{equation*}
we have the solution
\begin{equation*}
z(t) = \sum_{k = 0}^m q_k t^k + \sum_{k = 0}^{n+1} p_{k} t^k e^{-\tau t}
\end{equation*}
with 
\begin{alignat*}{1}
q_k =~& \sum_{w = k}^m d_w \frac{w!}{k!} \frac{(-1)^{w-k}}{\tau^{w-k+1}} \\
p_{k} =~& c_{k-1}/k, k \geq 1\\
p_0 =~& z(0) + \sum_{k=0}^m q_k (-1)^{k+1} \frac{k!}{\tau^{k+1}}.
\end{alignat*}
\end{lemma}

\begin{lemma} \label{lem_nonind}
Given the following equation for $z$, 
\begin{equation*}
\dot z = \sum_{k = 0}^m d_k t^k + \sum_{k = 0}^n c_{k} t^k e^{-\tau t},
\end{equation*}
we have the solution
\begin{equation*}
z(t) = \sum_{k = 0}^{m+1} q_k t^k + \sum_{k = 0}^{n} p_{k} t^k e^{-\tau t}
\end{equation*}
with 
\begin{alignat*}{1}
q_k =~& d_{k-1}/k, k \geq 1 \\
q_0 =~& z(0) + \sum_{k=0}^n d_k \frac{k!}{\tau^{k+1}} \\
p_{k} =~&-\sum_{w=k}^n c_w \frac{w!}{k!} \frac{1}{\tau^{w-k+1}}. 
\end{alignat*}
\end{lemma}

\begin{lemma}\label{lem_sumorder}
\begin{equation*}
\sum_{k=0}^m \sum_{w=0}^k f(k, w) = \sum_{k=0}^m \sum_{w=k}^m f(w,k)
\end{equation*}
\begin{equation*}
\sum_{k=0}^m \sum_{w=0}^{k-1} f(k, w) = \sum_{k=0}^{m-1} \sum_{w=k}^{m-1} f(w+1,k)
\end{equation*}
\begin{equation*}
\sum_{k=0}^m \sum_{w=0}^{k-2} f(k, w) = \sum_{k=0}^{m-2} \sum_{w=k}^{m-2} f(w,k)
\end{equation*}
\end{lemma}

\begin{lemma} \label{lem_Iter}
If $X_{a+1} = A_a X_a + B_a$, then
\[ X_a = X_1 \prod_{k=1}^{a-1} A_k + \sum_{k=1}^{a-1} B_k \prod_{w = k+1}^{a-1}A_w. \] 
\end{lemma}

\end{document}